\documentclass[10pt,article]{article}
\usepackage{makeidx} 
\usepackage{undertilde}
\usepackage{amsmath, amscd, amsthm, amssymb, mathrsfs,amsfonts,bm}
\usepackage{authblk}
\usepackage{mathrsfs}
\usepackage{graphicx}
\usepackage{color}

\newcommand{\ph}{\phantom}
\newcommand{\nn}{\nonumber}

\newcommand{\eps}{\epsilon}

\newcommand{\vareps}{\varepsilon}

\RequirePackage[dvipsnames]{xcolor}
\usepackage{color}

\usepackage{tikz,lipsum,lmodern}
\usepackage[most]{tcolorbox}
\usepackage{todonotes} % for comments

\usepackage[utf8]{inputenc}
\usepackage{amsmath}
\usepackage{amssymb}
\usepackage{graphicx}
\usepackage{caption}
\usepackage{subcaption}
\usepackage{float}
\usepackage{multirow}
\usepackage{url}
\usepackage{hyperref}
\usepackage{cleveref}
\usepackage{amsmath}
\usepackage{blindtext}
\usepackage{titlesec}
\usepackage{wasysym}

\usepackage{tikz}
\usetikzlibrary{shapes.geometric, arrows}
\definecolor{hawaiianblue}{rgb}{0.4,0.68,0.953}
\definecolor{sand}{rgb}{0.863,0.953,0.811}
\definecolor{tropicalgreen}{rgb}{0.251, 0.857,0.341}

\tikzstyle{blob} = [rectangle, rounded corners, minimum width=0.75cm, minimum height=0.75cm,text centered, draw=black, fill=hawaiianblue!30, align = center]

\tikzstyle{blob2} = [rectangle, rounded corners, minimum width=0.75cm, minimum height=0.75cm,text centered, draw=black, fill=tropicalgreen!47, align = center]

\tikzstyle{blob3} = [rectangle, rounded corners, minimum width=3cm, minimum height=1cm,text centered, draw=black, fill=sand!51, align = center]

\tikzstyle{blob4} = [rectangle, rounded corners, minimum width=3cm, minimum height=1cm,text centered, draw=black, fill=Dandelion, align = center]

\tikzstyle{blob5} = [rectangle, rounded corners, minimum width=3cm, minimum height=1cm,text centered, draw=black, fill=hawaiianblue!30, align = center]

\tikzstyle{arrow} = [thick,->,>=stealth, align = center]

\title{Hamiltonian formulation of gravity as a spontaneously-broken gauge theory of the Lorentz group 
}
\author{Mehraveh Nikjoo$^1$\footnote{\texttt{mehraveh.nikjoo@phdstud.ug.edu.pl}}\; and Tom Zlosnik$^1$\footnote{\texttt{thomas.zlosnik@ug.edu.pl}}
\\{\small \it $(1)$ Institute of Theoretical Physics and Astrophysics, University of Gda\'{n}sk, 80-308 Gda\'{n}sk, Poland}}

\begin{document}

\maketitle

\begin{abstract}
A number of approaches to gravitation have much in common with the gauge theories of the standard model of particle physics. In this paper, we develop the Hamiltonian formulation of a class of gravitational theories that may be regarded as spontaneously-broken gauge theories of the complexified Lorentz group $SO(1,3)_C$ with the gravitational field described entirely by a gauge field valued in the Lie algebra of $SO(1,3)_C$ and a `Higgs field' valued in the group's fundamental representation. The theories have one free parameter $\beta$  which appears in a similar role to the inverse of the Barbero-Immirzi parameter of Einstein-Cartan theory. However, contrary to that parameter, it is shown that the number of degrees of freedom crucially depends on the value of $\beta$. For non-zero values of $\beta$, it is shown the theories possesses three complex degrees of freedom, and for the specific values $\beta=\pm i$ an extension to General Relativity is recovered in a symmetry-broken regime. For the value $\beta=0$, the theory possesses no local degrees of freedom. A non-zero value of $\beta$ corresponds to the self-dual and anti-self-dual gauge fields appearing asymmetrically in the action, therefore in these models, the existence of gravitational degrees of freedom is tied to chiral asymmetry in the gravitational sector.
\end{abstract}

%\tableofcontents
\section{Introduction}
\label{introduction}
A great achievement of General Relativity has been the introduction of the notion of spacetime diffeomorphism symmetry as a cornerstone of gravitational physics. Less well known are formulations of non-gravitational physics which nonetheless possess the same symmetry - these theories are named \emph{parameterized} field theories. As an example, consider the action for degrees of freedom $q^{i}(\tau)$ in Newtonian mechanics:

\begin{align}
    S[q] &=  \int d\tau \bigg(\sum_{i} m_{i}\bigg(\frac{d}{d\tau}q^{i}\bigg)^{2}-V(q)\bigg) \label{newtonian_action}
\end{align}
Alternatively, one can consider an action where the Newtonian time $\tau$ is itself promoted to a dynamical field:

\begin{align}
\tau &\rightarrow \tau(\lambda)\\
    S[q,\tau] &= \int d\lambda \frac{d\tau}{d\lambda}\bigg(\sum_{i} m_{i}\bigg(\frac{d\tau}{d\lambda}\bigg)^{-2}\bigg(\frac{d}{d\lambda}q^{i}\bigg)^{2}-V(q) \bigg) \label{ppm_action}
\end{align}
Under a transformation generated by the infinitesimal vector $\zeta = \eps \partial_{\lambda} $:

\begin{align}
    \tau &\rightarrow \tau +{\cal L}_{\zeta}\tau ,\quad 
    q^{i} \rightarrow q^{i} + {\cal L}_{\zeta} q^{i}\label{qtautran}
\end{align}
- where ${\cal L}$ is the Lie derivative - 
the action (\ref{ppm_action}) changes by a boundary term and hence the transformations (\ref{qtautran}), which represent diffeomorphisms on the manifold coordinatized by $\lambda$, are a symmetry of the theory. This is a symmetry which is not present for the action (\ref{newtonian_action}), however the equations of motion following from (\ref{ppm_action}) admit the same solutions as those following from (\ref{newtonian_action}) if the gauge $\tau \overset{*}{=} \lambda$ is accessible, with the $\tau$ equation of motion expressing conservation of energy. The extension to parameterized field theory in higher dimensional special-relativistic actions is via the replacement of the Minkowski metric tensor $\eta_{\mu\nu}$ with

\begin{align}
    \eta_{\mu\nu} \rightarrow \eta_{IJ}\partial_{\mu}\phi^{I}(x)\partial_{\nu}\phi^{J}(x) \label{minkowski_to_dx}
\end{align}
where $\eta_{IJ}=\mathrm{diag}(-1,1,1,1)$ and $x^{\mu}$ are coordinates in spacetime. Analogously to the model (\ref{ppm_action}), actions with the replacement (\ref{minkowski_to_dx}) with the promotion of $\phi^{I}$ to dynamical fields then possess a four-dimensional spacetime diffeomorphism symmetry despite not including the gravitational interaction. If the gauge $\phi^{I} \overset{*}{=} x^{I}(x^{\mu})$ is accessible, where $x^{I}$ are fields playing the role of Minkowski coordinates in spacetime, then special-relativistic physics is recovered with the $\phi^{I}$ equations of motion corresponding to equations expressing conservation of stress energy.

It is possible then to recover a description of special-relativistic physics which nonetheless possesses the symmetries associated with gravitational theory. Can the gravitational interaction be recovered from this starting point and, if so, is the resulting theory General Relativity?
To take steps towards this, we note that actions built using (\ref{minkowski_to_dx}) have an additional symmetry which corresponds to:

\begin{align}
    \phi^{I} \rightarrow \Lambda^{I}_{\ph{I}J}\phi^{J}+P^{I} \label{global_poincare}
\end{align}
where $\Lambda^{I}_{\ph{I}J}\in SO(1,3)$ and $P^{I}$ are independent of coordinates $x^{\mu}$ and hence (\ref{global_poincare}) can be interpreted as a global Poincar\'{e} transformation acting on the fields $\phi^{I}$. 
If some of $\{\Lambda^{I}_{\ph{I}J},P^{I}\}$ do depend on position then the ordinary derivative $\partial_{\mu}\phi{^I}$ in (\ref{minkowski_to_dx}) no longer transforms homogeneously under the local generalization of (\ref{global_poincare}) and so actions containing (\ref{minkowski_to_dx}) will not be invariant under such transformations. This can be remedied by the introduction of fields $\{\omega^{I}_{\ph{I}J\mu}, \theta^{I}_{\mu}\}$ such that an operator ${\cal D}_{\mu}$ can be constructed, acting on $\phi^{I}$ as:

\begin{align}
    {\cal D}_{\mu}\phi^{I} \equiv \partial_{\mu}\phi^{I} + \omega^{I}_{\ph{I}J\mu}\phi^{J} + \theta^{I}_{\mu} \label{codiff}
\end{align}
It can be shown that (\ref{codiff}) transforms homogeneously under the local generalization of (\ref{global_poincare}) if 

\begin{align}
\omega^{I}_{\ph{I}J\mu} &\rightarrow \Lambda^{I}_{\ph{I}K}\omega^{K}_{\ph{K}L\mu}(\Lambda^{-1})^{L}_{\ph{L}J} - \partial_{\mu}\Lambda^{I}_{\ph{I}K}(\Lambda^{-1})^{K}_{\ph{K}J} \label{poin_spin_change}\\
\theta^{I}_{\mu} &\rightarrow \Lambda^{I}_{\ph{I}J}\theta^{J}_{\mu} - \partial_{\mu}P^{I} \label{poin_theta_change}
\end{align}
It follows then that the tensor

\begin{align}
    g_{\mu\nu} \equiv \eta_{IJ}{\cal D}_{\mu}\phi^{I}{\cal D}_{\mu}\phi^{J} \label{metric_tensor}
\end{align}
is invariant under the local Poincar\'{e} transformations and it is this composite object that will play the role of the metric tensor. Equation (\ref{metric_tensor}) can be seen as a definition of the metric tensor and is a composite object built from $\{\phi^{I},\omega^{I}_{\ph{I}J\mu},\theta^{I}_{\mu}\}$ which may be regarded as the fields describing gravity. Indeed, it is straightforward to build polynomial actions in these variables that correspond to the Einstein-Cartan formulation of gravity \cite{Grignani:1991nj}. One may additionally construct parameterized field theories corresponding to matter fields in fixed background metrics with non-vanishing curvature. Parameterized field theories which correspond to background metrics for De Sitter space and anti-De Sitter space have global symmetries and their gauging instead results in the Macdowell-Mansouri formulation of gravity. This is discussed in more detail in Appendix \ref{macdowellmansouri}.
Returning to the Minkowski case, remarkably, other theories of gravity may emerge if only a subgroup of the global Poincar\'{e} symmetry (\ref{global_poincare}) is promoted to a local one. If just the translational part is localized (hence gravity is described entirely by $\{\phi^{I},\theta^{I}_{\mu}\}$), then the resulting gravitational theory is teleparallel gravity \cite{Aldrovandi:2013wha}. On the other hand, one can consider the case where only the global Lorentz symmetry is promoted to a local one, hence gravity is to be described entirely by $\{\phi^{I},\omega^{I}_{\ph{I}J\mu}\}$. Remarkably, extensions of General Relativity can be recovered from the following family of actions: 

\begin{align}
S[\phi^{I},\omega^{IJ}_{\mu}] &= \frac{1}{2} \int d^{4}x\,\tilde{\eps}^{\mu\nu\alpha\beta}\big(\eps_{IJKL}+2\beta\eta_{K[I}\eta_{J]L}\big)D_{\mu}\phi^{I}  D_{\nu}\phi^{J}R^{KL}_{\ph{KL}\alpha\beta}(\omega) \label{khr_act_zero_one}
\end{align}
when $\beta = \pm i$ \cite{Zlosnik:2018qvg,Gallagher:2021tgx,Koivisto:2022uvd,Koivisto:2023epd}, where $D_{\mu}\phi^{I} = \partial_{\mu}\phi^{I} + \omega^{I}_{\ph{I}J\mu}\phi^{I}$ is the $SO(1,3)$-covariant derivative of $\phi^{I}$ and $R^{IJ}_{\ph{IJ}\alpha\beta}$ is the curvature two-form of the field $\omega^{I}_{\ph{I}J\mu}$. As such, the action (\ref{khr_act_zero_one}) is manifestly invariant under local $SO(1,3)$ transformations. As we are allowing for complex $\beta$, the action is potentially complex as well as the fields $(\phi^{I},\omega^{I}_{\ph{I}J\mu})$, with the local Lorentz symmetry being that of the \emph{complexified} Lorentz group $SO(1,3)_{C}$.
Specifically, an extension to General Relativity is recovered when $\phi^{I}\neq 0$ and with the metric tensor $g_{\mu\nu}$ identified with $\eta_{IJ}D_{\mu}\phi^{I}D_{\nu}\phi^{J}$; hence, formally this correspondence arises in a spontaneously-broken gauge theory described by gauge field $\omega^{I}_{\ph{I}J\mu}$ and Higgs field $\phi^{I}$, with non-vanishing values of the latter field breaking the symmetry $SO(1,3)_{C}$ to $SO(3)_{C}$ (for the case $\phi^{2}\equiv\eta_{IJ}\phi^{I}\phi^{J}<0$), $SO(1,2)_{C}$ ($\phi^{2}>0$), or $ISO(2)_{C}$ ($\phi^{2}=0)$ for non-vanishing $\phi^{I}$. The reader may note that upon replacement $D_{\mu}\phi^{I} \rightarrow e^{I}_{\mu}$ (where $e^{I}_{\mu}$ is the spacetime co-tetrad), the action is proportional to the Holst action \cite{Holst:1995pc}. The relation between these two actions is discussed in more detail in Section \ref{grav_acts}.

This paper aims to analyze the canonical structure of the action (\ref{khr_act_zero_one}) corresponding to general complex values of $\beta$. 
The canonical/Hamiltonian analysis is a very powerful tool to determine the dynamical structure of theories (see for example \cite{Blagojevic:2018dpz,Alexandrov:2019dxm,Blixt:2020ekl,Barker:2021oez,Alexandrov:2021qry,Sengupta:2022rbd,Karataeva:2022mll} for applications to different gravitational theories). It will be shown that the properties of the theory such as the ability to describe real spacetime metrics and even the number of degrees of freedom in the theory depend crucially on the value of $\beta$.

The structure of the paper is as follows: In Section \ref{math_prelims} we review several mathematical preliminaries necessary to describe the models under consideration here and their Hamiltonian form. In Section \ref{grav_acts} we briefly survey the action principles of some gravitational theories including the models that we will develop. In Section \ref{three_plus_one_lag} we construct the canonical formulation of the models in question considering the propagation of constraints, the classification of constraints, and the imposition of reality conditions on complex fields. In Section \ref{an_extension_of_gr} we show that some of the models correspond to an extension of General Relativity and in Section \ref{additional} we briefly consider matter couplings and possible additional terms in the gravitational action. In Section \ref{discussion} we discuss results and the potential for further study of these models.

\section{Mathematical Preliminaries}
\label{math_prelims}
We now review a number of mathematical preliminaries that will be helpful for the remainder of the analysis in this paper.

\subsection{Review of the 3+1 Decomposition}
\label{three_plus_one}

Maintaining general covariance is a typical requirement in gravitational theories. Dynamical fields are spacetime tensors and actions are coordinate-independent functionals built entirely from these fields. As such, there is a-priori no preferred notion of time in gravitational theory. However, significant insights can be gained into the structure of generally covariant theories by choosing a single coordinate by which to measure the change of fields over the spacetime manifold. This is the 3+1 decomposition of physical theories.

We assume that the spacetime manifold $M$ is topologically $R\times \Sigma$, where $\Sigma$ is a three-dimensional submanifold of $M$ and $R$ may be coordinatized by a number $t$ which can be thought of as `coordinate time'. However, we emphasize that $t$ is to be regarded as a number labeling different submanifolds of $M$ and may not be straightforwardly related to proper time as determined by a spacetime metric (indeed we will encounter solutions of the models (\ref{khr_act_zero_one}) where no notion of proper time exists). Furthermore, we are primarily interested in solutions to the theory in the bulk spacetime and will neglect boundary conditions and surface integrals. Hence our treatment will be exact only in cases where $\Sigma$ is a closed manifold \cite{Romano:1991up}.

Alongside the label $t$ for coordinate time, we will use a set of three coordinates $\{x^{a}\}$ ($a=1,2,3$) to cover the manifold $\Sigma$. As such, a vector $V$ and one-form $\sigma$ can be decomposed as follows:

\begin{align}
    V &= V^{\mu}\partial_{\mu} = V^{t}\partial_{t} + V^{a}\partial_{a}\\
    \sigma &= \sigma_{\mu}dx^{\mu} = \sigma_{t}dt + \sigma_{a}dx^{a} \label{three_plus_one_form}
\end{align}
By extension, if $M$ admits a metric tensor $g=g_{\mu\nu}dx^{\mu}\otimes dx^{\nu}$ then the following decomposition can be used:

\begin{align}
    g &= -N^{2}dt\otimes dt + q_{ab}(N^{a}dt + dx^{a})\otimes (N^{b}dt + dx^{b})\label{adm_metric}
\end{align}
Additionally, we will adopt the notation that for a function $f(t,x^{a})$ then $\dot{f}\equiv \partial_{t}f$ whilst $\partial_{a}f$ denotes the partial derivative with respect to a coordinate $x^{a}$.

\subsection{The Hamiltonian formalism}
\label{hamform}
A classical field theory will be described by an action $S$ which is a functional of fields $\chi^{{\cal A}}$ (which we use to denote a set of any tensor fields such as $V,\sigma,g$ as defined in Section \ref{three_plus_one}). The action can be written as an integral of a spacetime density $\tilde{\cal L}(\chi^{{\cal A}})$ called the Lagrangian density i.e.

\begin{align}
S[\chi^{{\cal A}}] &= \int \tilde{\cal L} dt d^{3}x
\end{align}
Given the 3+1 decomposition of tensorial fields, the Lagrangian density $\tilde{\cal L}$ can typically be written in the following form:

\begin{align}
\tilde{\cal L} &= \sum_{\cal B}a_{\cal B}(\chi^{\cal A},\partial_{a}\chi^{\cal A}) (\dot{\alpha}^{{\cal B}})^{2}+\sum_{\cal C}b_{\cal C}(\chi^{\cal A},\partial_{a}\chi^{\cal A}) \dot{\beta}^{{\cal C}} - \tilde{\cal U}(\chi^{\cal A},\partial_{a}\chi^{\cal A}) \label{three_plus_one_ell}
\end{align}
i.e. the collection of fields $\chi^{\cal A}$ can be divided into those which appear quadratically in time derivatives (the set $\{\alpha^{\cal B}\}$), linear in time derivatives (the set $\{\beta^{\cal C}\}$), and without time derivatives (the set which we will call $\{\gamma^{\cal D}\}$). By introducing auxiliary `velocity' fields ${\cal V}$ and Lagrange multiplier fields ${\cal P}$, the following \emph{extended} Lagrangian density can be constructed which yields identical equations of motion to (\ref{three_plus_one_ell}):

\begin{align}
    \tilde{\cal L} &=  \sum_{\cal B}\tilde{\cal P}_{\cal B}(\dot{\alpha}^{\cal B}-{\cal V}^{\cal B})+\sum_{\cal C}\tilde{\cal P}_{\cal C}(\dot{\beta}^{\cal C}-{\cal V}^{\cal C})+\sum_{\cal D}\tilde{\cal P}_{\cal D}(\dot{\gamma}^{\cal D}-{\cal V}^{\cal D}) \nn\\
    &+\sum_{\cal B}a_{\cal B}(\chi^{\cal A},\partial_{a}\chi^{\cal A})({\cal V}^{\cal B})^{2}+\sum_{\cal C}b_{\cal C}(\chi^{\cal A},\partial_{a}\chi^{\cal A}) {\cal V}^{\cal C} - \tilde{\cal U}(\chi^{\cal A},\partial_{a}\chi^{\cal A})
\end{align}
For the fields ${\cal V}^{\cal B}$, the equation of motion for ${\cal V}^{\cal B}$ allows for this field to be solved for in terms of the fields $(\chi^{\cal A},{\cal P}^{\cal B})$ allowing it to be eliminated from the variational principle. For fields ${\cal V}^{\cal C}$ and ${\cal V}^{\cal D}$, their equations of motion do not allow for the fields to be solved for and eliminated from the variational principle. The Lagrangian density then can be reduced to the following form:

\begin{align}
\tilde{\cal L} &=  \sum_{\cal A}\tilde{\cal P}_{\cal A}\dot{\chi}^{\cal A} - \tilde{\cal H}(\tilde{\cal P}_{\cal A},\chi^{\cal A},\partial_{a}\chi^{\cal A},{\cal V}^{\cal C},{\cal V}^{\cal D}) \label{hamilton}\\
\tilde{\cal H} &= {\cal H}_{0}(\tilde{\cal P}_{\cal A},\chi^{\cal A},\partial_{a}\chi^{\cal A})+ \sum_{{\cal C}}{\cal V}^{\cal C} {\cal C}^{\cal C}(\tilde{\cal P}_{\cal A},\chi^{\cal A},\partial_{a}\chi^{\cal A})+\sum_{{\cal D}}{\cal V}^{\cal D} {\cal C}^{\cal D}(\tilde{\cal P}_{\cal A},\chi^{\cal A},\partial_{a}\chi^{\cal A})
\end{align}
where the $({\cal V}^{C},{\cal V}^{D})$ are Lagrange multipliers (which enforce via their equations of motion $({\cal C}^{\cal C}=0,{\cal C}^{\cal D}=0)$) and $\tilde{\cal P}_{\cal A}$ consists of the collected fields $(\tilde{\cal P}_{\cal B},\tilde{\cal P}_{\cal C},\tilde{\cal P}_{\cal D})$. Equation (\ref{hamilton}) represents the Hamiltonian form of a theory, with stationarity of the action with respect to small variations of $(\chi^{\cal A},\tilde{\cal P}_{\cal A})$ yielding Hamilton's equations:

\begin{align}
\dot{\chi}^{\cal A} &= \{{\chi}^{\cal A} ,\int d^{3}x\tilde{\cal H}\}\label{ham_eq_1}\\
\dot{\tilde{\cal P}}_{\cal A} &= \{\tilde{\cal P}_{\cal A} ,\int d^{3}x\tilde{\cal H}\} \label{ham_eq_2}
\end{align}
where the Poisson bracket $\{{\cal F},{\cal G}\}$ between two functions ${\cal F}(\chi^{\cal A},\tilde{\cal P}_{\cal A})$ and ${\cal G}(\chi^{\cal A},\tilde{\cal P}_{\cal A})$ is defined to be:

\begin{align}
\{{\cal F},{\cal G}\} &\equiv  \int d^{3}x\sum_{\cal A}\bigg(\frac{\delta {\cal F}}{\delta \chi^{\cal A}}\frac{\delta {\cal G}}{\delta \tilde{\cal P}_{\cal A}}-\frac{\delta {\cal G}}{\delta \chi^{\cal A}}\frac{\delta {\cal F}}{\delta \tilde{\cal P}_{\cal A}}\bigg)
\end{align}
Furthermore, it follows from (\ref{ham_eq_1}) and (\ref{ham_eq_2}) that for some function ${\cal F}(\chi^{\cal A},\tilde{\cal P}_{\cal A})$ that

\begin{align}
\dot{\cal F} &= \{{\cal F},\int d^{3}x \tilde{\cal H}\} \label{time_deriv}
\end{align}
The equations of motion that follow from the variation of fields $({\cal V}^{\cal B},{\cal V}^{\cal C})$ are equations

\begin{align}
{\cal C}^{\cal B}(\tilde{\cal P}_{\cal A},\chi^{\cal A},\partial_{a}\chi^{\cal A}) &=0 ,\quad 
{\cal C}^{\cal C}(\tilde{\cal P}_{\cal A},\chi^{\cal A},\partial_{a}\chi^{\cal A}) =0
\end{align}
These equations represent constraints that the fields $(\chi^{\cal A},\tilde{\cal P}_{\cal A})$ must obey amongst themselves. If at some initial moment $t=t_{0}$, the constraints are satisfied, then it must further be required that the time derivative of these functions - defined via (\ref{time_deriv}) - is zero. This may imply additional constraints and, if so, their own time derivatives must be ensured to be zero. The process continues until no further constraints are generated. 

\subsection{Local Lorentz symmetry in gravitation and its complexification}
\label{lorentz_symmetry}
A slight modification to the variables describing gravity is necessary to couple gravity to fermionic fields. This requires the introduction of the co-tetrad field $e^{I}_{\mu}$ from which the metric $g_{\mu\nu}$ is constructed as

\begin{align}
g_{\mu\nu} = \eta_{IJ}e^{I}_{\mu}e^{J}_{\nu}\label{g_in_terms_of_e}
\end{align}
Where $\eta_{IJ}= \mathrm{diag}(-1,1,1,1)$. Due to the appearance of the matrix $\eta_{IJ}$, the expression (\ref{g_in_terms_of_e}) is invariant under transformations 

\begin{align}
e_{\mu}^{I} \rightarrow \Lambda^{I}_{\ph{I}J}e^{J}_{\mu}
\end{align}
where $\Lambda^{I}_{\ph{I}J} \in SO(1,3)$ i.e. $\Lambda^{I}_{\ph{I}J}$ are elements of the Lorentz group. The Weyl spinors of the standard model transform in the fundamental representations of the group $SL(2,C)$ and invariance under global $SL(2,C)$ transformations necessitates coupling to $e^{I}_{\mu}$ in spinor Lagrangians and the identification of $\Lambda^{I}_{\ph{I}J}$ as the $SO(1,3)$ element corresponding to that $SL(2,C)$ transformation. Note that (\ref{g_in_terms_of_e}) is invariant under transformations associated with $\Lambda^{I}_{\ph{I}J}$ which can depend on spacetime position. For spinorial actions then to be invariant under the associated \emph{local} $SL(2,C)$ transformation, it is necessary to introduce a field $\bar{\omega}^{I}_{\ph{I}J\mu}$ (where $\bar{\omega}^{IJ}_{\mu}=-\bar{\omega}^{JI}_{\mu}$ when an index has been raised with $\eta^{IJ}$, the matrix inverse of $\eta_{IJ}$) which transforms as a connection under local $SO(1,3)$ transformations (indeed, it should transform precisely as (\ref{poin_spin_change}) does). In General Relativity, this field is defined as the solution to the equation

\begin{align}
  \partial_{[\mu}e^{I}_{\nu]} + \bar{\omega}^{I}_{\ph{I}J[\mu}e^{J}_{\nu]} = 0
\end{align}
Therefore in General Relativity $\bar{\omega}^{I}_{\ph{I}J\mu}$ is determined by $e^{I}_{\mu}$ and its derivatives. A variation on General Relativity is provided by instead introducing a field $\omega^{I}_{\ph{I}J\mu}$ - called the spin connection - in place of $\bar{\omega}^{I}_{\ph{I}J\mu}(e,\partial e)$ which is to be regarded as an independent field with its own equations of motion. This is the Einstein-Cartan formulation of gravity. In its simplest form, the equation of motion for $\omega^{I}_{\ph{I}J\mu}$ yields a solution $\omega^{I}_{\ph{I}J\mu} = \bar{\omega}^{I}_{\ph{I}J\mu}(e,\partial e) + \dots$ where the dots denote terms linear in spinorial currents. 

A further generalization of the Einstein-Cartan model is provided by the Ashtekar chiral theory of gravity \cite{Ashtekar:1986yd}. To motivate this, we note that it has been up to now assumed that $\Lambda^{I}_{\ph{I}J}$ are elements of the real Lorentz group. However, the transformation (\ref{g_in_terms_of_e}) is invariant under $\Lambda^{I}_{\ph{I}J}$ belonging to the \emph{complexified} Lorentz group $SO(1,3)_{C}$ \footnote{Which in terms of properties of matrices $\Lambda^{I}_{\ph{I}J} \in SO(1,3)_{C}$ is defined to be the set of complex-valued matrices that satisfy $\eta_{IJ}=\eta_{KL}\Lambda^{K}_{\ph{K}I}\Lambda^{L}_{\ph{L}J}$ and $\mathrm{det}(\Lambda)=1$}. Can classical General Relativity also arise if the theory possesses a complex Lorentz symmetry? To understand the answer to this, it is first helpful to introduce self- and anti-self duality concepts for representations of $SO(1,3)_{C}$.

We have seen that it is helpful to introduce a field $e^{I}_{\mu}$ where under an $SO(1,3)$ transformation $e^{I}_{\mu}\rightarrow \Lambda^{I}_{\ph{I}J}e^{J}_{\mu}$. One can consider more general `Lorentz tensors' with a more complicated index structure. Particularly useful will be antisymmetric Lorentz tensors $F^{IJ}=-F^{JI}$ which transform as follows under Lorentz transformations

\begin{align}
F^{IJ} &\rightarrow \Lambda^{I}_{\ph{I}K}\Lambda^{J}_{\ph{J}L}F^{KL}
\end{align}
When the transformations are complexified Lorentz transformations, further decomposition of this (now complex-valued object) is possible. We can consider the following decomposition of $F^{IJ}$:

\begin{align}
F^{IJ}  &=  F^{+IJ} +  F^{-IJ}
\end{align}
where

\begin{align}
F^{\pm IJ} &=\frac{1}{2}(F^{IJ} \mp \frac{i}{2}\eps^{IJ}_{\ph{IJ}KL}F^{KL})\\
\frac{1}{2}\eps^{IJ}_{\ph{IJ}KL}F^{\pm KL} &= \pm i F^{\pm IJ} \label{sd_or_anti_sd}
\end{align}
where recall that $\eps_{IJKL}$ is the four-dimensional Levi-Civita symbol and indices are lowered or raised with $\eta_{IJ}$ and its matrix inverse $\eta^{IJ}$ respectively. If follows, for example, that for some matrix $Y_{IJ}$ that $Y_{IJ}F^{IJ\pm}= Y_{IJ}^{\pm}F^{\pm IJ}$. When the fields and Lorentz transformations are real then $F^{+IJ}$  and $F^{-IJ}$ are simply complex conjugates of one another. When the fields are complexified, they become genuinely independent objects. Equation (\ref{sd_or_anti_sd}) defines the property of self-dualness (here $F^{+IJ}$) or anti-self-dualness (here $F^{-IJ})$. It is possible to parameterize a self-dual or anti-self-dual Lorentz tensor in terms of a field $E^{I}$ as follows:

\begin{align}
    F^{\pm IJ} &= \frac{1}{2}(n^{[I}E^{J]} \mp \frac{i}{2}\eps^{IJKL}n_{K}E_{L})\\
    &=  (n^{[I}E^{J]})^{\pm} \label{nsd}
\end{align}
where $n^{I}$ is an arbitrary Lorentz vector of non-vanishing norm i.e. $\eta_{IJ}n^{I}n^{J}=\xi$ and $E_{I}n^{I}=0$, where $\xi <0$ for timelike $n^{I}$ and $\xi >0$ for spacelike $n^{I}$, and furthermore, for example, for a Lorentz tensor $W_{\dots [IJ]^{+}}$ which is self-dual in a pair of indices, we have:

\begin{align}
    W_{\dots [IJ]^{+}}F^{IJ^{+}} &= W_{\dots [IJ]^{+}}n^{I}E^{J} 
\end{align}
Finally, it is useful to define the following objects:

\begin{align}
{\cal K}^{\pm}_{IJKL} &= \frac{1}{2}(\eps_{IJKL} \pm 2i\eta_{I[K}\eta_{L]J}) \label{k_symbols}
\end{align}
where it can be shown that

\begin{align}
{\cal K}^{\pm}_{IJKL}F^{KL} &= \eps_{IJKL}F^{\pm KL}
\end{align}
i.e. the objects ${\cal K}^{\pm}_{IJKL}$ act to project out self- or anti-self-dual parts of an antisymmetric Lorentz tensor.

The spin connection $\omega^{IJ}_{\mu}=-\omega^{JI}_{\mu}$ present in Einstein-Cartan gravity is an antisymmetric Lorentz tensor and so can be decomposed into self-dual and anti-self-dual parts:

\begin{align}
\omega^{I}_{\ph{I}J\mu} = \omega^{+I}_{\ph{+I}J\mu}+\omega^{-I}_{\ph{-I}J\mu}
\end{align}
Upon complexification of the fields (which results from complexification of the $SO(1,3)$ gauge symmetry) then $(\omega^{+I}_{\ph{+I}J\mu},\omega^{-I}_{\ph{-I}J\mu})$ become truly independent fields 
and this independence will be shown in Section \ref{grav_acts} to be crucially important in the structure of gravitational fields based on this complexified Lorentz symmetry.

\subsection{Spacetime structure}
\label{spacetime_structure}
It will be very useful to relate some fields appearing in the canonical formalism to quantities appearing in the 3+1 metric formalism of gravity. To this end, we can use the following general parameterization of $e^{I}_{\mu}$ \cite{Peldan:1993hi}:

\begin{align}
    e^{I} = e^{I}_{t}dt +e^{I}_{a}dx^{a}
\end{align}
where 

\begin{align}
    e^{I}_{t} &= N N^{I} + N^{a}e^{I}_{a}
    \label{eit}\\
    q_{ab} &= \eta_{IJ}e^{I}_{a}e^{J}_{b}
\end{align}
where $N^{I}e_{Ia}=0$ and $N^{I}N_{I}=-1$. Computing the metric $g_{
\mu\nu} = \eta_{IJ}e^{I}_{\mu}e^{J}_{\nu}$ confirms that $(N,N^{a},q_{ab})$ should be identified with corresponding quantities appearing in (\ref{adm_metric}). Furthermore, it follows that

\begin{align}
\eta^{IJ} &= -N^{I}N^{J} + e^{aI}e_{a}^{J}
\end{align}
where  $e^{a}_{I} = q^{ab}e_{bI}$ where $q^{ab}$ is the matrix inverse of $q_{ab}$.
%Finally it's helpful to define $\eps_{IJK} = N^{L}\eps_{LIJK}$ hence
%$\eps_{IJKL} = -4 N_{[I}\eps_{JKL]}$. Hence:
%
%\begin{align}
%e^{a}_{I} &=\frac{1}%{2\sqrt{q}}\eps_{IJK}\tilde{\eps}^{abc}e^{J}_{b}e^{K}_{c}
%\end{align}
%

In the present work, the basic variables describing the gravitational field will be $(\phi^{I},\omega^{IJ}_{\mu})$ with the identification 

\begin{align}
    D_{\mu}\phi^{I} = \partial_{\mu}\phi^{I}+\omega^{I}_{\ph{I}J\mu}\phi^{J} = e^{I}_{\mu}
\end{align}
and hence we will look to identify the spacetime metric with $g_{\mu\nu}= \eta_{IJ}D_{\mu}\phi^{I}D_{\nu}\phi^{J}$. We may also usefully decompose $\phi^{I}$ into parts parallel with and orthogonal to $N^{I}$:

\begin{align}
\phi^{I} &=  \phi_{(N)} N^{I} + \varphi^{I}
\end{align}
where $\varphi_{I}N^{I}=0$. It follows then that $\phi_{I}e^{I}_{a} = \frac{1}{2}\partial_{a}\phi^{2} = \varphi_{I}e^{I}_{a}$ where we've used the fact that $N_{I}e^{I}_{a}$. Therefore, 

\begin{align}
    \varphi_{I} = \frac{1}{2}q^{ab}e_{Ia}\partial_{b}\phi^{2}
\end{align}
and hence

\begin{align}
\phi_{(N)} =  -\xi \sqrt{-\phi^{2}+ \frac{1}{4}q^{ab}\partial_{a}\phi^{2}\partial_{b}\phi^{2}}
\end{align}
where $\xi=\mp 1$. There are therefore two distinct options for the sign of $\phi_{(N)}$.

\section{Gravitational actions}
\label{grav_acts}
We now briefly survey several theories of gravitation and their symmetries. The action for Einstein's General Relativity can be written as:

\begin{align}
    S_{GR}[g_{\mu\nu}] &=  \frac{1}{16\pi G} \int_{M} d^{4}x\sqrt{-g} R+  \int_{\partial M} d^{3}y \,\tilde{\ell}_{GHY} 
\end{align}
Where the second term - the Gibbons-Hawking-York term - is a boundary action necessary to provide a well-defined variational principle. A spacetime diffeomorphism generated by a vector field $\xi^{\mu}$ transforms the spacetime metric as $g_{\mu\nu} \rightarrow g_{\mu\nu}+{\cal L}_{\xi}g_{\mu\nu}$ and it can readily shown that this changes the action by a boundary term and hence such diffeomorphisms are symmetries of the theory. 

As discussed in Section \ref{lorentz_symmetry}, the necessity to couple gravitation to fermions motivates the introduction of the fields $(e^{I}_{\mu},\omega^{IJ}_{\mu})$ as the descriptors of gravity. One of the simplest actions that can be constructed that has a General-Relativistic limit is the Einstein-Cartan Palatini action:

\begin{align}
S_{EC}[e^{I}_{\mu},\omega^{IJ}_{\mu}] &= \frac{1}{64\pi G} \int  d^{4}x\, \tilde{\eps}^{\mu\nu\alpha\beta}\eps_{IJKL}e^{I}_{\mu}e^{J}_{\nu}R^{KL}_{\ph{KL}\alpha\beta}(\omega) \label{palatini}
\end{align}
where $\tilde{\eps}^{\mu\nu\alpha\beta}$ is the Levi-Civita density and 

\begin{align}
R^{IJ}_{\ph{KL}\alpha\beta}(\omega) &= 2\partial_{[\mu}\omega^{IJ}_{\ph{IJ}\nu]} +  2\omega^{I}_{\ph{I}K[\mu}\omega^{KJ}_{\ph{KJ}\nu]}
\end{align}
are the components of the curvature tensor associated with $\omega^{I}_{\ph{I}J\mu}$. The Einstein-Cartan Palatini action possesses the same spacetime diffeomorphism symmetry as the action for General Relativity and is additionally invariant under local Lorentz transformations parameterized by matrices $\Lambda^{I}_{\ph{I}J}(x)$. As the spin connection is an antisymmetric tensor in its Lorentz indices, it decomposed into self- and anti-self-dual parts. Upon complexification of the local Lorentz symmetry, these two fields are in principle independent of one another and remarkably the equations of motion from the following actions

\begin{align}
S_{EC\pm} &= \frac{1}{32\pi G}\int  d^{4}x\,\tilde{\eps}^{\mu\nu\alpha\beta}{\cal K}^{\pm}_{IJKL}e^{I}_{\mu}e^{J}_{\nu}R^{KL}_{\ph{KL}\alpha\beta}(\omega)  = \frac{1}{32\pi G}\int d^{4}x \,\tilde{\eps}^{\mu\nu\alpha\beta}\eps_{IJKL}e^{I}_{\mu}e^{J}_{\nu}R^{KL\pm}_{\ph{\pm KL}\alpha\beta}(\omega) \nn\\
&= \frac{1}{32\pi G}\int d^{4}x\, \tilde{\eps}^{\mu\nu\alpha\beta}\eps_{IJKL}e^{I}_{\mu}e^{J}_{\nu}R^{KL\pm}_{\ph{\pm KL}\alpha\beta}(\omega^{\pm})   \label{ashtekar_lag}
\end{align}
yield the (complexified) Einstein's equations, where the solutions of real General Relativity can be imposed after the imposition of appropriate reality conditions on fields.
The actions (\ref{ashtekar_lag}) form Ashtekar's chiral formulation of gravity in which only one of the $\omega^{+IJ}_{\ph{+IJ}\mu}$ or $\omega^{-IJ}_{\ph{-IJ}\mu}$ appear in the action.

The models that we will look at are models where $g_{\mu\nu}$ is recovered from the combination (\ref{metric_tensor}) with $\theta^{I}_{\mu}=0$ and the dynamical variables of the theory will be $\{\phi^{I},\omega^{IJ}_{\ph{IJ}\mu}\}$. This suggests that ultimately we should identify $e^{I}_{\mu}$ as being recovered from the object $D_{\mu}\phi^{I}=\partial_{\mu}\phi^{I}+\omega^{I}_{\ph{I}J\mu}\phi^{J}$ and so, as in the case of Einstein-Cartan theory and Ashtekar's chiral theory we can look to construct Lagrangian densities which are quadratic in this field and linear in the curvature of $\omega^{IJ}_{\ph{IJ}\mu}$, anticipating that this may be the simplest action giving non-trivial gravitational dynamics \cite{Koivisto:2022uvd}:

\begin{align}
S &= \frac{\alpha}{2} \int d^{4}x\,\tilde{\eps}^{\mu\nu\alpha\beta}\big(\eps_{IJKL}+\frac{2}{\gamma}\eta_{K[I}\eta_{J]L}\big)D_{\mu}\phi^{I}  D_{\nu}\phi^{J}R^{KL}_{\ph{KL}\alpha\beta}(\omega) \label{khr_act_one}
\end{align}
where $\gamma=1/\beta$ and $\alpha$ is an overall multiplicative factor. With the aid of the symbols (\ref{k_symbols}) we can write (\ref{khr_act_one}) as:

\begin{align}
S[\phi^{I},\omega^{+IJ}_{\mu},\omega^{-IJ}_{\mu}] 
&= \int 
d^{4}x\,\tilde{\eps}^{\mu\nu\alpha\beta}\eps_{IJKL}D_{\mu}\phi^{I}  D_{\nu}\phi^{J}\bigg(g_{+}R^{KL}_{\ph{KL}\alpha\beta}(\omega^{+})  +g_{-} R^{KL}_{\ph{KL}\alpha\beta}(\omega^{-})\bigg)  \label{lag_kr_2} 
\end{align}
where

\begin{align}
	g_{\pm} &= \frac{\alpha}{2}\bigg(\frac{\gamma\mp i}{\gamma}\bigg) 
\end{align}
Equivalently, we can express the two free constants $(\alpha,\gamma)$ in (\ref{khr_act_one}) in terms of $(g_{+},g_{-})$ as:

\begin{align}
\alpha &= g_{+}+g_{-} \\
\gamma &= i\bigg(\frac{g_{+}+g_{-}}{g_{-}-g{+}}\bigg)
\end{align}
As in the case of (\ref{palatini}) and (\ref{ashtekar_lag}), we will find that the overall multiplicative constant $\alpha$ will be related to a multiple of the inverse of Newton's constant when a General-Relativistic limit of the model exists. We will henceforth set $\alpha=1$ and consider this as a choice of units and not a restriction on the space of theories.

By way of comparison the Palatini Lagrangian density (\ref{palatini}) of Einstein-Cartan theory can be generalized to:

\begin{align}
\tilde{\cal L}[e^{I}_{\mu},\omega^{\pm IJ}_{\mu}] &= \tilde{\eps}^{\mu\nu\alpha\beta}\eps_{IJKL}e_{\mu}^{I}e_{\nu}^{J}\bigg(\bar{g}_{+}R^{KL}_{\ph{KL}\alpha\beta}(\omega^{+}) + \bar{g}_{-}R^{KL}_{\ph{KL}\alpha\beta}(\omega^{-})\bigg) \label{genpal}
\end{align}
With the identification $\bar{g}_{\pm}=(\bar{\alpha}/2)(\bar{\gamma} \mp i)/\bar{\gamma}$ - with $\bar{\gamma}$ being the Barbero-Immirzi parameter and $\bar{\alpha}=1/(32\pi G)$- this is equal to the Holst Lagrangian density for gravity \cite{Holst:1995pc}. Formally the Lagrangian density (\ref{lag_kr_2}) can be recovered from (\ref{genpal}) via the replacement $e^{I}_{\mu} \rightarrow D_{\mu}\phi^{I}$\footnote{Interestingly, an approach based on an $SO(1,3)$ connection $\omega^{IJ}_{\mu}$, a field $\phi^{I}$, and an \emph{independent} field $e^{I}_{\mu}$ yields a variety of novel phenomenology \cite{Klinkhamer:2018mmk}.}. The Palatini action corresponds to $(\bar{g}_{+}= \bar{\alpha}/2,\bar{g}_{-}=\bar{\alpha}/2)$ ($\bar{\gamma}\rightarrow \infty)$ whilst the chiral Ashtekar theory corresponds to cases $(\bar{g}_{+}=1,\bar{g}_{-}=0)$ ($\bar{\gamma}=i)$ and $(\bar{g}_{+}=0,\bar{g}_{-}=1)$ ($\bar{\gamma}=-i)$ respectively. For general values of $\bar{\gamma}$, Einstein's theory can be recovered given suitable reality conditions placed on the fields  $(e^{I}_{\mu},\omega^{IJ}_{\mu} = \omega^{+IJ}_{\mu}+\omega^{-IJ}_{\mu})$. In contrast, we will find that the number of degrees of freedom for the models (\ref{lag_kr_2}) depends crucially on the values of $(g_{+},g_{-})$.
The value $\bar{\gamma}$ does affect the interaction between gravity and fermionic matter \cite{Perez:2005pm,Freidel:2005sn} and even in the absence of matter may be physically relevant upon quantization of (\ref{genpal}) \cite{Engle:2007wy}.

The focus of this paper will be to develop the canonical formulation of the action (\ref{lag_kr_2}).

\section{3+1 decomposition of Lagrangian density and canonical formulation}
\label{three_plus_one_lag}
We now proceed to perform the 3+1 decomposition of the Lagrangian density for the action (\ref{lag_kr_2}). Motivated by the 3+1 decomposition of a spacetime one-form introduced in (\ref{three_plus_one_form}), we introduce the following fields:

\begin{align}
\omega^{\pm IJ}_{\ph{\pm IJ}\mu}dx^{\mu} &=  \Omega^{\pm IJ}dt + \beta^{\pm IJ}_{\ph{\pm IJ}a} dx^{a} \label{omega_three_plus_one}
\end{align}
Furthermore, for notational compactness, we introduce the following quantities:

\begin{align}
R^{\pm IJ}_{ab} &\equiv 2\bigg(\partial_{[a}\beta^{\pm IJ}_{\ph{IJ}b]}+\beta^{\pm IK}_{\ph{CE}[a|}\beta_{ K\ph{B}|b]}^{\pm J}\bigg) \label{fdef}\\
e^{I}_{a} &\equiv \partial_{a}\phi^{I}+\beta^{I}_{\ph{I}Ja}\phi^{J} \label{edef}
\end{align}
where $\beta^{IJ}_{a} = \beta^{+IJ}_{a}+ \beta^{-IJ}_{a}$. Using the decomposition (\ref{omega_three_plus_one}) in (\ref{lag_kr_2}) we note that fields that appear with time derivatives (the fields $(\phi^{I},\beta^{\pm IJ}_{\ph{\pm IJ}a})$) appear linearly in those time derivatives. Recalling the discussion in Section \ref{hamform}, the extended Lagrangian density can be constructed by auxiliary introducing `velocity' fields - here $(V^{I},V_{a}^{\pm IJ})$ - which are constrained to be equal to the time derivatives of $(\phi^{I},\beta_{a}^{\pm IJ})$ on-shell, with this resulting as an equation of motion obtained by varying Lagrange multiplier fields $(\tilde{P}_{I},\tilde{P}^{\pm a}_{IJ})$. As the original time derivatives of fields appeared linearly, it follows that $(V^{I},V_{a}^{\pm IJ})$ equations of motion cannot be used to determine $(V^{I},V_{a}^{\pm IJ})$ and so these fields play the role of Lagrange multipliers ensuring constraints amongst the collected fields $(\phi^{I},\beta_{a}^{\pm IJ},\tilde{P}_{I},\tilde{P}^{\pm a}_{IJ})$. 
The remaining fields in the variational problem - $\Omega^{\pm IJ}$ - appear without time derivatives and appear linearly in the action. The extended Lagrangian density can be cast into the following form:

\begin{align}
\tilde{\cal L} &=\tilde{P}_{I}\dot{\phi}^{I}  +\tilde{P}^{+c}_{IJ}\dot{\beta}^{+ IJ}_{\ph{CD}c} +\tilde{P}_{IJ}^{-c}\dot{\beta}^{- IJ}_{\ph{IJ}c}-\tilde{\cal H} + \partial_{a}\tilde{\ell}^{a} \label{lag_kr_3}
\end{align}
Where

\begin{align}
\tilde{\cal H} &= -\Omega^{+IJ}\tilde{\cal G}^{+}_{IJ} - \Omega^{-IJ}\tilde{\cal G}^{-}_{IJ} + V^{I}\tilde{C}_{I} + V^{+IJ}_{a}\tilde{C}^{+a}_{IJ} + V^{-IJ}_{a}\tilde{C}^{-a}_{IJ} 
\end{align}

\begin{align}
	\tilde{\cal G}^{+}_{IJ} &\equiv D^{(\beta^{+})}_{c}\tilde{P}^{+c}_{IJ}+\big[\tilde{P}_{[I}\phi_{J]}\big]^{+} \\
	\tilde{\cal G}^{-}_{IJ} &\equiv D^{(\beta^{-})}_{c}\tilde{P}^{-c}_{IJ}+\big[\tilde{P}_{[I}\phi_{J]}\big]^{-}\\
	\tilde{C}_{I} &= \tilde{P}_{I} - 2g_{+} \eps_{IJKL}\tilde{\vareps}^{abc}e^{J}_{a}R_{bc}^{+KL} -2g_{-} \eps_{IJKL}\tilde{\vareps}^{abc}e^{J}_{a}R^{-KL}_{bc} \label{cicon}\\
	\tilde{C}^{+c}_{IJ} &= 	\tilde{P}_{IJ}^{+ c}-2g_{+} \big[\eps_{IJKL}\tilde{\vareps}^{abc}e^{K}_{a}e^{L}_{b}\big]^{+}\\
	\tilde{C}^{-c}_{IJ} &= \tilde{P}_{IJ}^{-c} - 2g_{-} \big[\eps_{IJKL}\tilde{\vareps}^{abc}e_{a}^{K}e_{b}^{L}\big]^{-}
\end{align}
where we've introduced the `spatial covariant derivative' with respect to any of ${\cal B}=\{\beta,\beta^{+},\beta^{-}\}$:

\begin{align}
D^{(\cal B)}_{a}Y^{AB\dots}_{\ph{AB\dots}CD\dots} &\equiv \partial_{a}  Y^{AB\dots}_{\ph{AB\dots}CD\dots} +  {\cal B}^{A}_{\ph{A}Ea}Y^{EB\dots}_{\ph{AB\dots}CD\dots} 
+  {\cal B}^{B}_{\ph{A}Ea}Y^{AE\dots}_{\ph{AB\dots}CD\dots}  + \dots\nn\\
& - {\cal B}^{E}_{\ph{E}Ca} Y^{AB\dots}_{\ph{AB\dots}ED\dots} - {\cal B}^{E}_{\ph{E}Da} Y^{AB\dots}_{\ph{AB\dots}CE\dots} + \dots
\end{align}
The total derivative $\partial_{a}\tilde{\ell}^{a}$ will be neglected as an ignorable boundary term. Recall that the fields $(e^{I}_{a},R^{\pm IJ}_{ab})$ are themselves composed of $(\phi^{I},\beta^{\pm IJ}_{a})$ and their spatial derivatives and so the Lagrangian density (\ref{lag_kr_3}) can be interpreted as an action in canonical form with a phase space coordinatized by canonical pairs $(\phi^{I},\tilde{P}_{I})$ and $(\beta^{\pm IJ}_{a},\tilde{P}^{\pm a}_{IJ})$ with a primary Hamiltonian density $\tilde{\cal H}$ comprised entirely of constraints $(\tilde{\cal G}^{\pm}_{IJ},\tilde{\cal C}_{I},\tilde{\cal C}^{\pm a}_{IJ})$ enforced by stationarity of the action with respect to small variations of Lagrange multiplier fields $(\Omega^{\pm IJ},V^{I},V^{\pm IJ}_{a})$. 
\subsection{The evolution of constraints}
For functions $F$ and $G$ depending on phase space variables we can define the Poisson bracket as follows:

\begin{align}
\{F,G\} \equiv \int d^{3}x \bigg[\frac{\delta F}{\delta \beta^{+IJ}_{d}}\frac{\delta G}{\delta \tilde{P}^{+d}_{IJ}}+\frac{\delta F}{\delta \beta^{-IJ}_{d}}\frac{\delta G}{\delta \tilde{P}^{-d}_{IJ}}+\frac{\delta F}{\delta \phi^{I}}\frac{\delta G}{\delta \tilde{P}_{I}}\bigg] - (F\leftrightarrow G) \label{poisson_bracket}
\end{align}
Explicit forms for functional derivatives with respect to the phase space fields are given in Appendix \ref{functional_derivatives}.
Time evolution of fields $(\phi^{I},\beta_{a}^{\pm IJ},\tilde{P}_{I},\tilde{P}^{\pm a}_{IJ})$ are obtained from the Euler-Lagrange equations following from variation of (\ref{lag_kr_3}). Therefore for a function $F$ of these fields then

\begin{align}
\dot{F} &= \{F,\int d^{3}x \tilde{\cal H} \} \label{time_ev}
\end{align}
Finally, it will be useful to introduce the notion of smearing of phase space functions. For a phase space function $F$, its smearing with a test function $\xi(x)$ (i.e. a function which does not depend on the phase space fields) is defined as:

\begin{align}
F[\xi] &\equiv \int d^{3}x \xi F
\end{align}
We require that time evolution according to (\ref{time_ev}) preserves the set of constraints $(\tilde{\cal G}^{\pm}_{IJ},\tilde{\cal C}_{I},\tilde{\cal C}^{\pm a}_{IJ})$. For illustrative purposes, a detailed example of the evaluation of the Poisson bracket of two constraints is presented in Appendix \ref{poisson_example}. It turns out that preservation of ${\cal G}^{\pm}_{IJ}$ under time evolution is ensured if the primary constraints are satisfied - indeed these constraints generate self and anti-self dual Lorentz transformations, and hence the Poisson bracket of these constraints with any other constraint (when both constraints are smeared) will be proportional to constraints up to boundary terms that we do not consider in this analysis. For the remaining constraints $(\tilde{\cal C}_{I},\tilde{\cal C}^{\pm a}_{IJ})$ we recover the following equations:

\begin{align}
\partial_{t} \tilde{C}_{I} &\approx   -V^{+KL}_{a}W^{a+}_{KLI}-V^{-KL}_{a}W^{a-}_{KLI}
\label{prop1}\\
\partial_{t}\tilde{C}_{IJ}^{+e} &\approx 
\bigg[g_{+}Y^{de}_{IJKL}+g_{+}Y^{de}_{KLIJ}\bigg]^{+}V_{d}^{+KL}+ \bigg[g_{+}Y^{de}_{IJKL}+g_{-}Y^{de}_{KLIJ}\bigg]^{+}V_{d}^{-KL}\nn\\
& +W^{e+}_{IJM}V^{M}\label{prop2}\\
\partial_{t}\tilde{C}_{IJ}^{-e} & \approx
\bigg[g_{-}Y^{de}_{IJKL}+g_{-}Y^{de}_{KLIJ}\bigg]^{-}V_{d}^{-KL}+ \bigg[g_{-}Y^{de}_{IJKL}+g_{+}Y^{de}_{KLIJ}\bigg]^{-}V^{+KL}_{d}\nn\\
%g_{-}\bigg[(\bm{V}_{d}^{-KL} +\bm{V}_{d}^{+KL})Y^{de}_{IJKL}\bigg]^{-}+\bigg[(g_{-}\bm{V}_{d}^{-KL} +g_{+}\bm{V}_{d}^{+KL})Y^{de}_{KLIJ}\bigg]^{-}\nn\\
&+W^{e-}_{IJM}V^{M}\label{prop3}
\end{align}
where $\approx$ denotes weak equality (i.e. equality up to the addition of constraints)
and where we've defined the following:

\begin{align}
Y^{de}_{IJKL} &= Y^{de}_{[IJ][KL]} \equiv  4\tilde{\vareps}^{dbe}\eps_{MIJ[K}\phi_{L]}e^{M}_{b}  \label{yten}\\
W^{e \pm }_{IJK} &\equiv \bigg[2(g_{\mp}-g_{\pm})\tilde{\vareps}^{ebc}\eps_{KMN[I}\phi_{J]}R^{\mp MN}_{bc}+ g_{\pm}\tilde{\vareps}^{ebc}\phi_{K}\eps_{IJMN}R^{\pm MN}_{bc}\bigg]^{\pm}
\end{align}
This object is a tensor in the space coordinatized by antisymmetric Lorentz tensors. 
Now, adding the projection of (\ref{prop2}) along $V^{+IJ}_{e}$ to the projection of (\ref{prop3}) along $V^{-IJ}_{e}$ we obtain:

\begin{align}
\label{consistency}
V^{+IJ}_{e}\partial_{t}\tilde{C}^{+e}_{IJ}+V^{-IJ}_{e}\partial_{t}\tilde{C}^{-e}_{IJ} &= -V^{I}\partial_{t}\tilde{C}_{I}
\end{align}
It is shown in Appendix \ref{vint} that $V^{I}$ contains information about the lapse ($N$) and shift ($N^{a}$) functions from the 3+1 decomposition of the spacetime metric (\ref{adm_metric}) and whose functional forms are arbitrary insofar as they reflect the freedom to foliate spacetime in different ways. As such, (\ref{consistency}) can be taken to imply that generally the preservation of constraints $\tilde{C}^{\pm a}_{IJ}$ under time evolution implies the preservation of $\tilde{C}_{I}$. 

To proceed, it will be useful to explicitly work out self-dual and anti-self dual projections of indices of the object (\ref{yten}), for which explicit expressions are given in equation (\ref{Yplusmin}) in Appendix \ref{usefulresults}. Using the decomposition of self and anti-self-dual Lorentz tensors defined in (\ref{nsd}) and making use of the vector $N^{I}$ introduced in (\ref{eit}) we can define the objects ${\cal V}^{\pm I}_{a}$ as follows

\begin{align}
    V_{a}^{\pm IJ} &= \big[N^{[I}{\cal V}^{\pm J]}_{a}\big]^{\pm}
\end{align}
where ${\cal V}^{\pm I}_{a} N_{I} = 0$. We would like to find out whether the constraint propagation equations (\ref{prop2}) and (\ref{prop3}) for vanishing left-hand side amount to equations which uniquely determine ${\cal V}^{\pm I}_{a}$
Projecting these equations along $N^{I}$ we have:

\begin{align}
0&\approx \ph{-}ig_{+}\tilde{\vareps}^{dbe}\frac{1}{2}\partial_{b}\phi^{2}\eta_{IJ} {\cal V}_{d}^{+J} +(g_{+}-g_{-})N^{L}N^{J}Y^{de}_{[LI]^{+}[JK]^{-}}{\cal V}_{d}^{-K}+W^{e+}_{KIJ}N^{K}V^{J}\label{prop_sim_1}\\
0&\approx -ig_{-} \tilde{\vareps}^{dbe}\frac{1}{2}\partial_{b}\phi^{2} \eta_{IJ} {\cal V}_{d}^{-J} +(g_{-}-g_{+})N^{L}N^{J}Y^{de}_{[LI]^{-}[JK]^{+}}{\cal V}_{d}^{+K}+W^{e-}_{KIJ}N^{K}V^{J} \label{prop_sim_2}
\end{align}
where we've used the fact that $\phi_{K}e^{K}_{b}=\frac{1}{2}\partial_{b}\phi^{2}$.

\subsubsection{The special case $g_{+}=g_{-}=1/2$}
A number of terms in (\ref{prop_sim_1}) and (\ref{prop_sim_2}) vanish when $g_{+}=g_{-}$, substantially simplifying the equations. If we further define

\begin{align}
    R^{\pm IJ} &= \big[N^{[I}{\cal R}_{bc}^{\pm J]}\big]^{\pm}%= \frac{1}{2}\bigg(n^{[I}{\cal R}_{bc}^{\pm J]}\mp\frac{i}{2}\eps^{IJ}_{\ph{IJ}KL}n^{K}{\cal R}^{\pm L}_{bc}\bigg)
\end{align}
where $N_{I}{\cal R}^{\pm I}_{bc}=0$ then (\ref{prop_sim_1}) and (\ref{prop_sim_2}) take the form:

\begin{align}
0&\approx - \tilde{\vareps}^{dbe}\partial_{b}\phi^{2} \delta_{IJ} {\cal V}_{d}^{+J} +\tilde{\vareps}^{ebc}{\cal R}^{+}_{Ibc} \phi_{J}V^{J}\label{prop_sim_1}\\
0&\approx -\tilde{\vareps}^{dbe}\partial_{b}\phi^{2} \delta_{IJ} {\cal V}_{d}^{-J}  +\tilde{\vareps}^{ebc}{\cal R}^{-}_{Ibc}\phi_{J}V^{J} \label{prop_sim_2}
\end{align}
where $\delta_{IJ} = \eta_{IJ}+ N_{I}N_{J}$ and recall that $\phi^{2}\equiv \phi_{I}\phi^{I}$. Equations (\ref{prop_sim_1}) and (\ref{prop_sim_2}) can be regarded as a pair of linear inhomogeneous equations, involving either ${\cal V}^{+I}_{a}$ or ${\cal V}^{-I}_{a}$, each of which can be regarded as a 9 dimensional vector. The quantity $M^{de}_{\ph{de}IJ}=\tilde{\vareps}^{dbe}\partial_{b}\phi^{2} \delta_{IJ}$ that multiplies these vectors in each equation can be thought of as a $9\times 9$ matrix. If this matrix is invertible then equations (\ref{prop_sim_1}) and (\ref{prop_sim_2}) uniquely determine ${\cal V}^{\pm I}_{a}$. However, the matrix has the following three null-eigenvectors

\begin{align}
    \partial_{a}\phi^{2} S^{I(i)}
\end{align}
where $i=1,2,3$ and $S^{I(i)}N_{I}=0$. This suggests that the matrix $M^{de}_{\ph{de}IJ}$ is not invertible and not all ${\cal V}^{\pm I}_{a}$ can be determined from these equations. Acting on (\ref{prop_sim_1}) and (\ref{prop_sim_2}) with these null eigenvectors we obtain

\begin{align}
0 = \tilde{\vareps}^{ebc}R^{\pm IJ}_{bc}\partial_{e}\phi^{2} \label{nonseccon}
\end{align}
However, these are not new constraints on the phase space. This is due to the following identity:

\begin{align}
D^{(\beta^{\pm})}_{c}\tilde{C}^{\pm c}_{IJ} &\approx \tilde{\cal G}^{\pm}_{IJ}+ig_{\pm}\tilde{\eps}^{bca}R^{\pm}_{IJ ca}\partial_{b}\phi^{2}
\end{align}
i.e. the primary constraints imply (\ref{nonseccon}). The existence of null eigenvectors of $M^{ab}_{\ph{ab}IJ}$ shows that not all components of ${\cal V}^{\pm I}_{a}$ are determined by the propagation constraint equations; however, some can be solved for. Acting on the propagation equations with $\utilde{\eps}_{fea}\partial^{a}\phi^{2}\eta^{KI}$  (where $\partial^{a}\phi^{2}\equiv q^{ab}\partial_{b}\phi^{2}$) and introducing the projector:
${\cal P}^{a}_{\ph{a}b}  = \delta^{a}_{b} - \frac{\partial_{b}\phi^{2}\partial^{a}\phi^{2}}{(\partial_{c}\phi^{2}\partial^{c}\phi^{2})}$, the equations can be solved to yield:

\begin{align}
\bar{\cal V}^{+I}_{a} &\equiv {\cal P}^{b}_{\ph{d}a} {\cal V}_{b}^{ +I} \approx \frac{2}{(\partial_{b}\phi^{2}\partial^{b}\phi^{2})}\partial^{c}\phi^{2}\phi_{J}{\cal R}^{+ I}_{ca}V^{J}\nn\\
\bar{\cal V}^{-I}_{a} &\equiv {\cal P}^{b}_{\ph{d}a} {\cal V}_{b}^{ -I} \approx \frac{2}{(\partial_{b}\phi^{2}\partial^{b}\phi^{2})}\partial^{c}\phi^{2}\phi_{J}{\cal R}^{- I}_{ca}V^{J}
\end{align}
where we have introduced the convention that barred Lagrange multipliers denote multipliers that have been solved for in terms of fields in phase space. It will further be useful to introduce the symbols 

\begin{align}
\bar{V}^{\pm IJ}_{a} &= \big[N^{[I}{\cal V}^{\pm J]}_{a}\big]^{\pm} = \bar{U}^{\pm IJ}_{\ph{\pm IJ} Ka}V^{K}
\end{align}
where by inspection $\bar{U}^{\pm IJ}_{\ph{\pm IJ} Ka}=\frac{2}{(\partial_{b}\phi^{2}\partial^{b}\phi^{2})}\partial^{c}\phi^{2}\phi_{K}R^{\pm IJ}_{\ph{\pm IJ}ca}$.

\subsubsection{General case}
We now look at the case where $g_{+}\neq g_{-}$.
It can be seen that in the general case, all components of the ${\cal V}^{\pm I}_{a}$ can be solved for if the following $9\times 9$ complex matrix is invertible:

\begin{align}
Y^{de}_{[KI]^{(1)+}[LJ]^{(2)-}}N^{K}N^{L}&=\frac{1}{2}\tilde{\vareps}^{dbe}(e^{L}_{b}\phi^{M}+
 e_{b}^{M}\phi^{L})
N_{M}N^{K}\eps_{IJKL} -\frac{i}{2}\tilde{\vareps}^{dbe}e^{K}_{b}
(\phi_{L}N^{L}N_{K}\delta_{IJ}-\phi_{J}\delta_{IK})\nn\\
&-\frac{i}{2}\tilde{\vareps}^{dbe}e^{K}_{b}((\phi_{L}N^{L}N_{K} + \phi_{K})\delta_{IJ}-\phi_{I}\delta_{JK}) \label{ypmin}
\end{align}
Calculation shows that the determinant of the matrix (\ref{ypmin}) is generally non-zero if $\mathrm{det}[q_{ab}]\neq 0$ and $\phi_{I}\phi^{I}\neq 0$. These quantities can generally be assumed to be non-zero, hence in the case $g_{+}\neq g_{-}$ the matrix inverse exists, and all components of ${\cal V}^{\pm I}_{a}$ can be determined from the propagation of primary constraints in terms of $(R^{+IJ}_{ab},R^{-IJ}_{ab},\phi^{I})$. We will again use the convention that barred Lagrange multipliers denote their form when expressed in terms of these phase space fields and their spatial derivatives. It will further be useful to introduce the symbols 

\begin{align}
\bar{V}^{\pm IJ}_{a} &= \bar{Z}^{\pm IJ}_{\ph{\pm IJ} Ka}V^{K}
\end{align}
This expresses the formal solution for $\bar{V}^{\pm IJ}_{a}$ but we note that a closed expression for $\bar{Z}^{\pm IJ}_{\ph{\pm IJ} Ka}$ for general values of $(g_{+},g_{-})$ may be difficult to obtain. This completes the constraint analysis in the general case as no more constraints have been generated. We can now determine Hamilton's equations of motion, which for any values of $(g_{+},g_{-})$ are given by:

\begin{align}
\dot{\phi}^{I} &= V^{I} - (\Omega^{+IJ} + \Omega^{-IJ})\phi_{J} \label{phidot}\\
\dot{\beta}^{+IJ}_{c} &\approx D^{(\beta^{+})}_{c}\Omega^{+IJ} + V^{+IJ}_{c} \label{betaplusdot}\\
\dot{\beta}^{-IJ}_{c} &\approx D^{(\beta^{-})}_{c}\Omega^{-IJ} + V^{-IJ}_{c} \label{betaminusdot}\\
\dot{\tilde{P}}_{K} &\approx D^{(\beta)}_{a}\leftmoon^{a}_{K}+(\Omega^{+I}_{\ph{+I}K}+ \Omega^{-I}_{\ph{-I}K})\tilde{P}_{I}\\
\dot{\tilde{P}}^{+a}_{KL} &\approx -4D_{b}^{(\beta^{+})}(V^{I}g_{+}\eps_{IJKL}\tilde{\eps}^{eba}e^{J}_{e})^{+}- (2\Omega^{+\ph{L}J}_{\ph{+}[L}\tilde{P}^{+a}_{K]J} +\leftmoon^{a}_{[K}\phi_{L]})^{+}\\
\dot{\tilde{P}}^{-a}_{KL} &\approx  -4D_{b}^{(\beta^{-})}(V^{I}g_{-}\eps_{IJKL}\tilde{\eps}^{eba}e^{J}_{e})^{-}- (2\Omega^{-\ph{L}J}_{\ph{-}[L}\tilde{P}^{-a}_{K]J} +\leftmoon^{a}_{[K}\phi_{L]})^{-}
\end{align}
where we have defined

\begin{align}
\leftmoon_{K}^{a} &= 2\big(V^{I}(g_{+}R^{+JP}_{bc}+g_{-}R^{-JP}_{bc})+2(V^{+IJ}_{b}g_{+}+V^{-IJ}_{b}g_{-})e^{P}_{c}\big)\eps_{IJKP}\tilde{\eps}^{abc}
\end{align}
Weak equalities have been used to allow for the fact that the constraint propagation analysis fixes some or all of $V^{\pm IJ}_{a}$ (depending on whether $g_{+}=g_{-}$ or not) to explicitly depend on phase space fields. Hence, additional terms involving derivatives of $V^{\pm IJ}_{a}$ with respect to these fields will appear in Hamilton's equations but they will all be proportional to constraints $\tilde{C}^{\pm a}_{IJ}$ and so vanish on the constraint surface.

We finally point out an exotic solution that has not been covered by the prior analysis: that in which $\phi^{I}=0$ throughout spacetime. From (\ref{phidot}) we see that if $\phi^{I}=0$ initially then it will remain so only if $V^{I}=0$. From the results of Appendix \ref{vint} we see that this implies that the function $N=0$ and furthermore if $\phi^{I}=0$ then $q_{ab}=0$ and hence from equation (\ref{adm_metric}) the spacetime metric $g_{\mu\nu}=0$. Furthermore, if $\phi^{I}=0$ then $Y^{de}_{IJKL}$ and $W^{e\pm}_{IJK}$ are zero and hence $V^{\pm KL}$ are completely undetermined by the constraint propagation equations, thus implying from (\ref{betaplusdot}) and (\ref{betaminusdot}) that the time evolution of fields $\beta^{\pm IJ}_{a}$ is undetermined. It is unclear whether such solutions play any phenomenological role.
\subsection{The Algebra of Constraints}
\label{algebraofconstraints}
Having completed the calculation of propagation of constraints, we can now classify the primary constraints in terms of whether they are first-class constraints (i.e. their Poisson bracket with all other constraints weakly vanishes) or second-class constraints (i.e. their Poisson bracket with some constraints does not weakly vanish). The character of the constraints depends on the values of $(g_{+},g_{-})$. Given a phase space with dimensionality $P$ per spatial point and a number of first-class constraints $F$ and second class constraints $S$ per spatial point, the number of degrees of freedom per spatial point is defined to be:

\begin{align}
    DOF = \frac{1}{2}(P-2F-S)
\end{align}
This number ultimately represents how many numbers per spatial point at an initial time $t_{0}$ may be freely chosen as to generate a unique, physically distinctive solution (as opposed to a solution that is equal to another one up to a gauge transformation) to the equations of motion. As the theory is generally complex, this may be a complex number - and we will refer to this number as the number of complex degrees of freedom that the model possesses. The choice of complex numbers within the initial data of fields may result in complex fields in spacetime such as a complex spacetime metric $g_{\mu\nu}$. The physical interpretation of an imaginary component to the spacetime metric is not clear and so one path is to look to determine what choice of initial data ensures the reality of fields such as the metric tensor. This question is considered in detail in Sections \ref{realityconditions} and \ref{minkowksigravwaves}.

\subsubsection{Case $g_{+}\neq g_{-}$}
\begin{figure}[h!]
\begin{center}

%% Nodes 
\begin{tikzpicture}[node distance=1.2cm]
\node (cicon) [blob2] {$\tilde{C}^{I}$};
\node(gplus)[blob, right of = cicon, xshift = 1.0cm]{$\tilde{\cal{G}}^{+IJ}$};
\node(gminus)[blob, right of = gplus, xshift = 1.0cm]{$\tilde{\cal{G}}^{-IJ}$};
\node(cplus)[blob2, right of = gminus, xshift = 1.0cm]{$\tilde{C}^{+a}_{IJ}$};
\node(cminus)[blob2, right of = cplus, xshift = 1.0cm]{$\tilde{C}^{-a}_{IJ}$};
\node(hamcon)[blob, below of = cicon,xshift=4.5cm, yshift = -1.0cm]{$\tilde{\mathscr{H}}^{I}=\tilde{C}^{I}+ \bar{Z}^{+ JKI}_{\ph{+JKI} a}\tilde{C}^{+a}_{JK}+\bar{Z}^{- JKI}_{\ph{-JKI}a}\tilde{C}^{-a}_{JK}$};

%% Arrows

\draw [dotted,arrow] (cicon) --node[anchor = south]{}(hamcon);
\draw [dotted,arrow] (cplus) --node[anchor = south]{}(hamcon);
\draw [dotted,arrow] (cminus) --node[anchor = south]{}(hamcon);
%\draw [arrow] (aw) --node[anchor = east]{Constrain  $V$} (cec);
%\draw[arrow](cec) --node[anchor = north]{Solve for $\omega(e)$}(gr);

\end{tikzpicture}

\end{center}

\caption{The structure of constraints in the case $g_{+}\neq g_{-}$. First-class constraints are blue whilst constraints that are individually second-class are shown as green. The constraint analysis reveals that a linear combination of second-class constraints yields the first-class constraints $\tilde{\mathscr{H}}^{I}$.}
\label{general_constraints}
\end{figure}
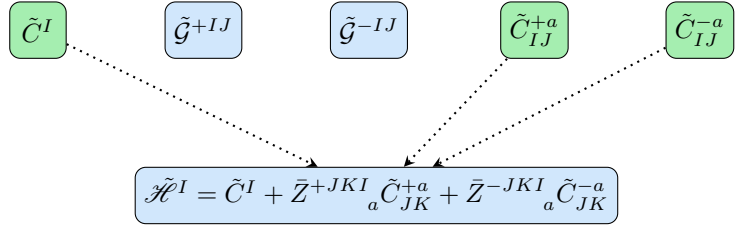

For the general case $g_{+}\neq g_{-}$, the classification of constraints is illustrated in Figure \ref{general_constraints}. Given the classification of constraints, we can now count how many complex degrees of freedom the model possesses. The dimensionality of the phase space per spatial point is
\begin{equation*}
    P = 8\, (\phi^{I},\tilde{P}_{I})+ 18 \,(\beta^{+IJ}_{a},\tilde{P}^{+a}_{IJ})+18\, (\beta^{-IJ}_{a},\tilde{P}^{-a}_{IJ}) = 44
\end{equation*}
 The number of first-class constraints is 
 \begin{equation*}
     F= 3\,(\tilde{G}^{+IJ})+3\,(\tilde{G}^{-IJ})+4\,(\tilde{\mathscr{H}}^{I})=10
 \end{equation*}
and the number of second-class constraints is
\begin{equation*}
    S= 9\,(\tilde{C}^{+ a}_{IJ})+9\,(\tilde{C}^{- a}_{IJ})=18
\end{equation*}
The number of degrees of freedom per spatial point is therefore 

\begin{align}
DOF &= \frac{1}{2}(P-2F-S)\nn\\
 &= 3
\end{align}
%
%% Flow diagram

\subsubsection{Case $g_{+}= g_{-}$}
\begin{figure}[h!]
%Left-right symmetric case
\begin{center}

%% Nodes 
\begin{tikzpicture}[node distance=1.2cm]
\node (cicon) [blob2] {$\tilde{C}^{I}$};
\node(gplus)[blob, right of = cicon, xshift = 1.0cm]{$\tilde{\cal{G}}^{+IJ}$};
\node(gminus)[blob, right of = gplus, xshift = 1.0cm]{$\tilde{\cal{G}}^{-IJ}$};
\node(cplusproj)[blob2, right of = gminus, xshift = 1.0cm]{${\cal P}^{a}_{\ph{a}b}\tilde{C}^{+b}_{IJ}$};
\node(cplussym)[blob, right of = cplusproj,xshift=1.0cm]{$\partial_{a}\phi^{2}\tilde{C}^{+a}_{IJ}$};
\node(cminusproj)[blob2, right of = cplussym, xshift = 1.0cm]{${\cal P}^{a}_{\ph{a}b}\tilde{C}^{-b}_{IJ}$};
\node(cminussym)[blob, right of = cminusproj,xshift=1.0cm]{$\partial_{a}\phi^{2}\tilde{C}^{-a}_{IJ}$};
\node(hamcon)[blob, below of = cicon,xshift=6.58cm, yshift = -1.0cm]{$\tilde{\mathscr{H}}^{I}=\tilde{C}^{I}+ \bar{U}^{+ JKI}_{\ph{+JKI} a}\tilde{C}^{+a}_{JK}+\bar{U}^{- JKI}_{\ph{-JKI}a}\tilde{C}^{-a}_{JK}$};

%% Arrows

\draw [dotted,arrow] (cicon) --node[anchor = south]{}(hamcon);
\draw [dotted,arrow] (cplusproj) --node[anchor = south]{}(hamcon);
\draw [dotted,arrow] (cminusproj) --node[anchor = south]{}(hamcon);
%\draw [arrow] (aw) --node[anchor = east]{Constrain  $V$} (cec);
%\draw[arrow](cec) --node[anchor = north]{Solve for $\omega(e)$}(gr);

\end{tikzpicture}

\end{center}

\caption{The structure of constraints in the case $g_{+}= g_{-}$. First class constraints are blue whilst second class constraints are green. Unlike in the case $g_{+}\neq g_{-}$, 
a subset of the individual $\tilde{C}^{\pm a}_{IJ}$ constraints are first class. As in the $g_{+}\neq g_{-}$ constraint analysis reveals that a linear combination of individually second class constraints yields the first class constraints $\tilde{\mathscr{H}}^{I}$.}
\label{specific_constraints}
\end{figure}
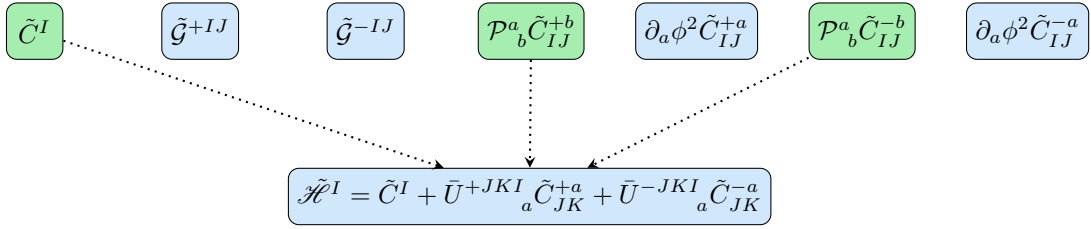

For the special case $g_{+}=g_{-}$, the classification of constraints is illustrated in Figure \ref{specific_constraints}. As in the previous case, the dimensionality of the phase space per spatial point is
\begin{equation*}
    P = 8\, (\phi^{I},\tilde{P}_{I})+ 18 \,(\beta^{+IJ}_{a},\tilde{P}^{+a}_{IJ})+ 18\, (\beta^{-IJ}_{a},\tilde{P}^{-a}_{IJ}) = 44
\end{equation*}
However, now the number of first-class constraints is 
\begin{equation*}
    F= 3\,(\tilde{G}^{+IJ})+3\,(\tilde{G}^{-IJ})+4\,(\tilde{\mathscr{H}}^{I})+3(\partial_{a}\phi^{2}\tilde{C}^{+ a}_{IJ})+3\,(\partial_{a}\phi^{2}\tilde{C}^{- a}_{IJ})=16
\end{equation*}
and the number of second-class constraints is
\begin{equation*}
    S= 6\,({\cal P}^{a}_{\ph{a}b}\tilde{C}^{+ b}_{IJ})+6\,({\cal P}^{a}_{\ph{a}b}\tilde{C}^{- b}_{IJ})=12
\end{equation*}
The number of degrees of freedom per spatial point is therefore 

\begin{align}
DOF &= \frac{1}{2}(P-2F-S)\nn\\
 &= 0
\end{align}

Therefore the theory with $g_{+}=g_{-}$ possesses no degrees of freedom. Indeed, this case has more first-class constraints than the case $g_{+} \neq g_{-}$ and so it is to be expected that this specific case has more symmetry than the general case. The precise additional symmetry that the theory possesses compared to the case $g_{+}\neq g_{-}$ can be demonstrated in the Lagrangian formalism. It is useful to write the action for general $(g_{+},g_{-})$ in the language of differential forms:

\begin{align}
    S &= 2\int \eps_{IJKL} D\phi^{I} \wedge D\phi^{J}\wedge \big(g_{+}R^{+KL}+g_{-}R^{-KL}\big)
    \label{form_action}
\end{align}
where we use $D$ to denote the covariant derivative $d+\omega$ according to the entire spin connection $\omega=  \omega^{+}+\omega^{-}$. Now consider the following field transformation:

\begin{align}
\phi^{I} \rightarrow \phi^{I} ,\quad 
\omega^{IJ\pm}_{\mu} \rightarrow \omega^{IJ\pm}_{\mu}+ \partial_{\mu}\phi^{2} \xi^{\pm IJ}  \label{field_transformation}
\end{align}
Under (\ref{field_transformation}) the action changes as:

\begin{align}
\delta S &= \int 4(g_{+}-g_{-})\eps_{IJKL}d\phi^{2} \wedge D\phi^{J}  \wedge \bigg(\xi^{-IM}R^{+KL}+\xi^{+IM}R^{-KL}\bigg)\phi_{M} \\
&+2D\bigg(\eps_{IJKL}d\phi^{2}  \wedge D\phi^{I} \wedge D\phi^{J}(g_{+}\xi^{+KL}+g_{-}\xi^{-KL})\bigg)
\end{align}
Therefore in the case $g_{+}=g_{-}$ and only in this case does the action change by a total derivative - and hence boundary term - under the transformation of fields (\ref{field_transformation}). This result holds even `off-shell' and therefore the field transformation is a symmetry of the theory \cite{Harlow:2019yfa}. Note that the transformation (\ref{field_transformation}) when applied to the pullback of $\omega^{\pm IJ}_{\mu}$ to surfaces of constant time $\beta^{\pm IJ}_{a}$ agrees with the transformation generated by the first class constraints $\partial_{a}\phi^{2} \tilde{C}^{\pm a}_{IJ}$ i.e. 

\begin{align}
    \delta \beta^{\pm IJ}_{a} &= \{\beta^{\pm IJ}_{a}, \partial_{b}\phi^{2}\tilde{C}^{\pm b}_{KL}[\xi^{\pm KL}]\} \nn\\
    &= \partial_{a}\phi^{2}\xi^{\pm IJ}
\end{align}
\subsection{Reality conditions}
\label{realityconditions}
We have seen that models with $g_{+}\neq g_{-}$ possess three complex degrees of freedom according to the definition of Section \ref{algebraofconstraints}. Because of the inherent complexity of the theory, in principle, the Hamiltonian may generate classical time evolution of fields to become complex even if initially real at some moment $t=t_{0}$. A standard requirement is that the spacetime metric be real. From (\ref{adm_metric}) this is ensured if fields $(N,N^{a},q_{ab})$ are real. From Appendix \ref{vint} it's clear that $(N,N^{a})$ are real if $V^{I}$ and $e^{I}_{a}$ are real \footnote{The complex-valued fields that coordinatize the phase space may combine with complex $V^{I}$ to produce real four-dimensional metrics of Euclidean signature but we do not explore that possibility in this work}. We will require that $V^{I}$ is real and that  $q_{ab}=\eta_{IJ}e^{I}_{a}e^{J}_{b}$ be initially real and for this realness to be preserved by time evolution. Additionally, anticipating that the norm $\phi^{2}=\phi_{I}\phi^{I}$ will have physical significance, this quantity should also be required to be real. Time evolution is generated by the Hamiltonian

\begin{align}
H = -\tilde{\cal G}^{+}_{IJ}[\Omega^{+IJ}] - \tilde{\cal G}^{-}_{IJ}[\Omega^{-IJ}] + \tilde{\cal C}_{I}[V^{I}] + \tilde{\cal C}^{+d}_{IJ}[V_{d}^{+IJ}] +\tilde{\cal C}^{-d}_{IJ}[ V_{d}^{-IJ}] \label{the_hamiltonian}
\end{align}
Then, recalling the definition (\ref{edef}) of $e^{I}_{b}$ we have for the general case $g_{+}\neq g_{-}$ that

\begin{align}
\partial_{t}q_{ab} &= 2\eta_{IJ}e^{I}_{a}\{\partial_{b}\phi^{J}+\beta^{J}_{\ph{J}Kb}\phi^{K},H\}\nn\\
&=2\eta_{IJ}e^{I}_{a}\big(-\partial_{b}V^{J}+(\bar{Z}^{+JKL}_{b}+ \bar{Z}^{-JKL}_{b})V_{L}\phi_{K} +\beta^{J}_{\ph{J}Kb} V^{K}  \big)\\
\partial_{t}\phi^{2} &= 2\phi_{I}\{\phi^{I},H\}\nn\\
&=2\phi_{I}V^{I} \label{realphisq}
\end{align}
From (\ref{realphisq}), we see that given our assumption that $V^{I}$ is real then an initially real $\phi^{2}$ remains real if $\phi_{I}$ is real.
In the general case, it's likely not possible to find closed expressions for $(\bar{Z}^{+JKL}_{b},\bar{Z}^{-JKL}_{b})$. However, they will depend on the generally complex $(g_{+},g_{-})$ which may create an imaginary part to $\partial_{t}q_{ab}$ even if initial data for the phase space fields is real. It is challenging to determine in the case of general $(g_{+},g_{-})$ whether maintaining the reality of $q_{ab}$ generates further constraints on the complex phase space. We will see, however, that in the special cases $(g_{+}=1,g_{-}=0)$ and $(g_{+}=0,g_{-}=1)$ that contact with familiar results from the Ashtekar model is possible. First, it is helpful to illustrate a manifestation of the challenge of finding reality conditions in the general case in a simple physical example: the propagation of linear metric perturbations on a Minkowski space background.

\subsection{The propagation of metric perturbations on Minkowski space}
\label{minkowksigravwaves}
The Euler-Lagrange equations following from the Lagrangian density (\ref{khr_act_one}) have a solution $R^{IJ}_{\ph{IJ}\mu\nu}(\omega)=0$ \cite{Koivisto:2022uvd}. 
Thus a gauge can be found where $\omega^{IJ}_{\mu}\overset{*}{=}0$ and
$g_{\mu\nu} \overset{*}{=}\eta_{IJ}\partial_{\mu}\phi^{I}\partial_{\nu}\phi^{J}$.
If $R^{IJ}_{\ph{IJ}\mu\nu}(\omega)=0$ then it turns out that $\phi^{I}$ is otherwise undetermined by the field equations and hence $\phi^{I}$ can take a profile where it forms a set of Minkowski coordinate such that $g_{\mu\nu}=\eta_{\mu\nu}$ i.e. Minkowski space is a solution to the theory for general values of $\gamma$.
Now we restrict ourselves to a wedge where $\phi^{2} =\eta_{IJ}\phi^{I}\phi^{J}<0$ and adopt a Lorentz gauge where $\bar{\phi}^{I} \overset{*}{=} T\delta^{I}_{0}$. Then, using $T$ as a time coordinate and denoting $x^{a}$ as spatial coordinates on the surface of constant time we have

\begin{align}
ds^{2} &=- dT \otimes dT+ T^{2}\delta_{ij}E^{i}_{a}E^{j}_{b}dx^{a}\otimes dx^{b}
\end{align}
where $i,j=1,2,3$ index spatial coordinates, $E^{1}_{a}dx^{a}=d\chi, E^{2}_{a}dx^{a} = \sinh(\chi)d\theta, E^{3}_{a}dx^{a}=\sinh(\chi)\sin(\theta)d\phi$.
Therefore we can identify the region $\phi_{I}\phi^{I}<0$ in Minkowski spacetime coordinatized by $(t,x^{a})$ as an open Friedmann-Robertson-Walker universe with scale factor $a=T$. This is a Milne wedge and without loss of generality, we choose the upper Milne wedge of Minkowski spacetime. Now consider the following small perturbations to the spin connection

\begin{align}
    \delta \omega^{0i} &= \frac{1}{2}H^{i}_{\ph{i}j}E^{j}\\
    \delta \omega^{ij} &= a^{ij}dt+\eps^{ijl} W_{l}^{\ph{g}k}E_{k}
\end{align}
where $(H^{ij},a^{ij},W^{ij})$ are symmetric and traceless. These perturbations produce a perturbed spatial metric:

\begin{align}
    \delta g_{ab} = H^{jk}E_{j a}E_{k b}
\end{align}
It can be shown that the perturbation $(a_{ij},W_{ij})$ can be solved for algebraically in the field equations in terms of first derivatives of $H_{ij}$ and so eliminated from the variational principle to recover the following Lagrangian density  

\begin{align}
   \frac{1}{2} \tilde{\cal L}_{(H)} &=a^{3}\bigg(-\frac{1}{\gamma^{2}}\dot{H}_{ij}\dot{H}^{ij}-\frac{1}{a^{2}}h^{ab} {\cal D}_{a}{ H}^{ij}{\cal D}_{b}{H}_{ij}-\frac{2k}{a^{2}} H_{ij}H^{ij}\bigg) \label{grav_wave_prop}
\end{align}
where here $h^{ab}$ is the unperturbed inverse spatial metric on the surface of constant $T$ - with spatial curvature constant $k=-1$, $a=T$ - is the `cosmological' scale factor, 
and where $H_{ij}$ is assumed to have support in the upper Milne wedge. In the case $\gamma^{2}=-1$ (which corresponds to either $(g_{+}=1,g_{-}=0)$ or $(g_{+}=0,g_{-}=1)$) the Lagrangian density reduces to that of General Relativity, describing the lightlike propagation of the spin-2 perturbation $H_{ij}$ on this background. For all other values of $\gamma$, the propagation of $H_{ij}$ is at a different speed, and it is readily seen that if $\gamma^{2}$ has an imaginary component then an initially real perturbation $H_{ij}(t=t_{0})$ will evolve by the equations of motion to become complex and hence in this simple case the only way to preserve reality of the spatial metric would be to constrain $H_{ij}(t=t_{0})=0$, showing that the reality conditions can reduce the number of propagating degrees of freedom.

\section{An extension of General Relativity}
\label{an_extension_of_gr}
We see from (\ref{grav_wave_prop}) that only for the case $\gamma^{2}=-1$ does the propagation of gravitational waves correspond to that of General Relativity and we will now focus on this case. The condition $\gamma^{2}=-1$ encompasses two independent possibilities: $(g_{+}=1,g_{-}=0)$ and $(g_{+}=0,g_{-}=1)$. We will first focus on the former case and later demonstrate how the latter case can be straightforwardly recovered. When $(g_{+}=1,g_{-}=0)$, the primary Hamiltonian density simplifies considerably. Recalling its general form

\begin{align}
\tilde{\cal H} &= -\Omega^{+IJ}\tilde{\cal G}^{+}_{IJ} - \Omega^{-IJ}\tilde{\cal G}^{-}_{IJ} + V^{I}\tilde{C}_{I} + V^{+IJ}_{a}\tilde{C}^{+a}_{IJ} + V^{-IJ}_{a}\tilde{C}^{-a}_{IJ} ,
\end{align}
the constraints now simplify to:

\begin{align}
	\tilde{\cal G}^{+}_{IJ} &\equiv D^{(\beta^{+})}_{c}\tilde{P}^{+c}_{IJ}+\big[\tilde{P}_{[I}\phi_{J]}\big]^{+} \\
	\tilde{\cal G}^{-}_{IJ} &\equiv D^{(\beta^{-})}_{c}\tilde{P}^{-c}_{IJ}+\big[\tilde{P}_{[I}\phi_{J]}\big]^{-}\\
	\tilde{C}_{I} &= \tilde{P}_{I} - 2\eps_{IJKL}\tilde{\vareps}^{abc}e^{J}_{a}R_{bc}^{+KL}\\
	\tilde{C}^{+c}_{IJ} &= 	\tilde{P}_{IJ}^{+ c}-2\big[\eps_{IJKL}\tilde{\vareps}^{abc}e^{K}_{a}e^{L}_{b}\big]^{+}\\
	\tilde{C}^{-c}_{IJ} &= \tilde{P}_{IJ}^{-c}  \label{self_dual_cminus}
\end{align}
Given these simplifications, it is possible to algebraically solve for the $(\beta^{-IJ}_{a},\tilde{P}^{-a}_{IJ})$ and eliminate them from the variational problem. Firstly, from (\ref{self_dual_cminus}) we have $\tilde{P}_{IJ}^{-a}=0$. Then, recalling the definition (\ref{edef}), the constraint $\tilde{C}^{+c}_{IJ}$ can be regarded as an equation for which one can solve for $\beta^{-IJ}_{a}=\beta^{-IJ}_{a}(\beta^{+},\tilde{P}^{+},\phi,\partial\phi)$. Therefore the second-class constraints can be solved. Given these solutions, the constraint $\tilde{\cal G}^{-}_{IJ}$ simplifies to $\big[\tilde{P}_{[I}\phi_{J]}\big]^{-}=0$; if we decompose $\tilde{P}^{I}=  \tilde{\Pi}\phi^{I} + \tilde{P}_{\perp}^{I}$, where   $\tilde{P}_{\perp}^{I}\phi_{I}=0$, then $\tilde{\cal G}^{-}_{IJ}=0$ can be taken to imply the solution $\tilde{P}_{\perp}^{I}=0$ which we now adopt.
Additionally the quantity $V^{I}\tilde{C}_{I}$ can be expressed in terms of quantities $(\utilde{N}\equiv N/\sqrt{q},N^{a})$ and the remaining phase space variables using a number of useful results detailed in Appendix \ref{usefulresults}, so that the gravitational Lagrangian density can be written:

\begin{align}
\tilde{\cal L}[\tilde{P}^{+a}_{IJ},\beta^{+IJ}_{a},\tilde{\Pi},\phi^{2},\Omega^{+IJ},\utilde{N},N^{a}] &=\frac{1}{2}\tilde{\Pi}\dot{\phi^{2}} +\tilde{P}^{+a}_{IJ}\dot{\beta}^{+ IJ}_{\ph{CD}a}+\Omega^{+IJ}\bigg(D^{(\beta^{+})}_{a}\tilde{P}^{+a}_{IJ}\bigg) \nn\\
&- \utilde{N}\bigg(\xi  \tilde{\Pi}\sqrt{q}\sqrt{-\phi^{2}+ \frac{1}{4}q^{ab}\partial_{a}\phi^{2}\partial_{b}\phi^{2}} -\frac{1}{4}\tilde{P}^{+a}_{IK}\tilde{P}^{+bK}_{\ph{+cK}J}R^{+IJ}_{\ph{+IJ}ab}\bigg)\nn\\
&- N^{a}\bigg( \frac{1}{2}\tilde{\Pi}\partial_{a}\phi^{2}+\tilde{P}^{+b}_{IJ}F^{+IJ}_{ab}-\beta^{+ IJ}_{a} D^{(\beta^{+})}_{b}\tilde{P}^{+b}_{IJ}\bigg) \label{lag_den_fin}
\end{align}
where $\sqrt{q}$ and $q^{ab}$  can be expressed in terms of $\tilde{P}^{+a}_{IJ}$ using (\ref{momdet}) and we have redefined $\Omega^{+IJ}\rightarrow \Omega^{+IJ}+N^{a}\beta_{a}^{+IJ}$ so that the constraint obtained from the $N^{a}$ equation of motion when smeared with a field $\zeta^{a}$ generates (non-Lorentz covariant) spatial diffeomorphisms $f\rightarrow f + {\cal L}_{\zeta}f$ on fields $f$ in the phase space coordinatized by $(\phi^{2},\tilde{\Pi},\beta^{+IJ}_{a},\tilde{P}^{+a}_{IJ})$. The Lagrangian density that we have recovered corresponds to the canonical formulation of Ashtekar's theory of gravity \cite{Romano:1991up} coupled to a field $\phi^{2}=\eta_{IJ}\phi^{I}\phi^{J}$ whose dynamics is classically that of a pressureless perfect fluid when $\phi^{2}<0$ \cite{Husain:2011tk,Koivisto:2022uvd}. If $\phi^{2}<0$ and local coordinates are selected so that $\partial_{a}\phi^{2}$ for some region of spacetime then the energy density of the fluid is of sign $\xi \tilde{\Pi}$ and so the choice of the sign of $\xi = \mp 1$ reflects the relative sign of the energy density and $\tilde{\Pi}$. More generally, $\phi^{2}$ by definition is not positive-definite and its equation of motion is:

\begin{align}
\frac{1}{2}\partial_{t}\phi^{2} &=  N\xi  \sqrt{-\phi^{2}+ \frac{1}{4}q^{ab}\partial_{a}\phi^{2}\partial_{b}\phi^{2}} +\frac{1}{2}N^{a}\partial_{a}\phi^{2}\label{dphi_sq}
\end{align}
The equation of motion for $\phi^{2}$ differs from the equation of motion for, for example, the Higgs boson of the standard model; unlike that case, $\partial_{t}\phi^{2}$ is independent of the field's momentum $\tilde{\Pi}$ and generally the right-hand side of (\ref{dphi_sq}) will be non-zero whenever the spacetime metric is non-degenerate meaning that generally the magnitude of $\phi^{2}$ will vary throughout spacetime. Notably there do exist solutions to the theory's field equations where, furthermore, the sign of $\phi^{2}$ varies throughout spacetime \cite{Koivisto:2022uvd}. 

If instead, we had chosen the parameters $(g_{+}=0,g_{-}=1)$ we would have instead recovered the \emph{anti-}self-dual formulation of Ashtekar's theory coupled to an effective matter component described by $(\phi^{2},\tilde{\Pi})$, the Lagrangian density of which can be recovered from (\ref{lag_den_fin}) by the replacement of $(\beta^{+IJ}_{a},\tilde{P}^{+a}_{IJ})$ with $(\beta^{-IJ}_{a},\tilde{P}^{-a}_{IJ})$.
\subsection{The field $\phi^{2}$ as a standard of time}
\label{phisqtime}
Can the degree of freedom $\phi^{2}=\eta_{IJ}\phi^{I}\phi^{J}$ act as a useful `clock field' in physics? If we choose initial data such that $\partial_{a}\phi^{2}|_{t=t_{0}}=0$ then the equation of motion for $\phi^{2}$ at the initial moment becomes:

\begin{align}
\frac{1}{2}\partial_{t}\phi^{2} &=  \xi N\sqrt{-\phi^{2}} \label{phisqev}
\end{align}
 The condition that $\partial_{a}\phi=0$ will be preserved if $N=N(t)$. Given this condition, we can regard the integration of equation (\ref{phisqev}) as providing a functional relation between $\phi^{2}$ and $t$. If $N=1/\xi$ (which, recall from (\ref{adm_metric}) implies that $t$ corresponds to proper time) then $\phi^{2}=-t^{2}$ is a solution. If instead $N=1/(2\xi\sqrt{t})$ then $\phi^{2}=-t$ is a solution. Therefore we will associate a choice of $N(t)$ with a partial spacetime gauge fixing ($N^{a}$ is left undetermined) and denote the associated time coordinate as $t_{\phi}$. We may additionally solve the Hamiltonian constraint for $\tilde{\Pi}$ in this gauge to yield the following Lagrangian density:

\begin{align}
\tilde{\cal L}[\tilde{P}^{+a}_{IJ},\beta^{+IJ}_{a},\Omega^{+IJ},N^{a}] &\overset{*}{=}\tilde{P}^{+a}_{IJ}\frac{\partial }{\partial t_{\phi}}{\beta}^{+ IJ}_{\ph{CD}a}-{\cal H}_{Phys}\nn\\
&+\Omega^{+IJ}\bigg(D^{(\beta^{+})}_{a}\tilde{P}^{+a}_{IJ}\bigg) - N^{a}\bigg( \tilde{P}^{+b}_{IJ}F^{+IJ}_{ab}-\beta^{+ IJ}_{a} D^{(\beta^{+})}_{b}\tilde{P}^{+b}_{IJ}\bigg) \label{lag_den_fin_gf}\\
{\cal H}_{Phys} &= -\frac{N(t_{\phi})}{4\sqrt{q}}\tilde{P}^{+a}_{IK}\tilde{P}^{+bK}_{\ph{+cK}J}R^{+IJ}_{\ph{+IJ}ab} 
\end{align}
Therefore we recover an action principle with a phase space coordinatized by fields $(\tilde{P}^{+a}_{IJ},\beta_{a}^{+IJ})$, with a physical Hamiltonian density ${\cal H}_{Phys}$ and phase space constraints implemented by stationarity of the action under variation of $(\Omega^{+IJ},N^{a})$.
Note that now $N$ appears not as an independent field but as a fixed function of time and only in gauges $\partial_{t_{\phi}}N(t_{\phi})=0$ does the physical Hamiltonian not have explicit time-dependence (as opposed to the implicit time-dependence it possesses via its dependence on time-varying phase space fields), which corresponds to the case where $\sqrt{-\phi^{2}}$ measures metric proper time. The extension to General Relativity is encoded in the fact that ${\cal H}_{Phys}$ is not constrained to vanish. Non-vanishing values would be interpreted as an effective energy density of an additional dust-like gravitating component in physics \cite{Koivisto:2022uvd}. 

It is instructive to determine the reality conditions on fields evolving according to the equations of motion that follow from (\ref{lag_den_fin_gf}). We can require the spacetime metric to be initially real. This implies that $N$ should be real, and we can furthermore impose that $N^{a}$ is real throughout spacetime. Additionally, the spatial metric $q_{ab}$ should be required to be initially real, and its reality should be preserved by time evolution. Equivalently, we can require that the densitized inverse spatial metric remain real, which from (\ref{momdet}) amounts to requiring that $\tilde{P}^{+c}_{IJ}\tilde{P}^{+dIJ}$ remains real. From Hamilton's equations, we have that

\begin{align}
\frac{\partial}{\partial t_{\phi}}\tilde{P}^{a}_{IJ} &= -\frac{N(t_{\phi})}{2}D_{b}^{(\beta^{+})}\bigg(\frac{1}{\sqrt{q}}\tilde{P}_{[I|K}^{[b} \tilde{P}^{a]K}_{\ph{a]K}|J]}\bigg)  + 2D_{b}\bigg(N^{[b}\tilde{P}^{a]}_{IJ}\bigg)
\end{align}
and hence given the assumed reality of $(N,N^{a},q_{ab})$ at the initial time, it follows that the following term must be real so that $\tilde{P}^{+(a}_{IJ}\tilde{P}^{b)IJ}$ remains real:

\begin{align} 
\tilde{P}^{+(a|IJ}D_{c}^{(\beta^{+})}\bigg(\frac{1}{\sqrt{q}}\bigg(\tilde{P}_{IK}^{+c} \tilde{P}^{+|b)K}_{\ph{+ aK}J}-\tilde{P}_{IK}^{+|b)} \tilde{P}^{+cK}_{\ph{+ bK}J}\bigg)\bigg) 
\end{align}
This condition is equivalent to the corresponding condition on the momenta of the self-dual spin connection $\beta^{+IJ}_{a}$ in Ashtekar's chiral formulation of General Relativity \cite{Ashtekar:1991hf}.

The presence of degrees of freedom which would classically correspond to a pressureless perfect fluid have been proposed as solutions to the `problem of time' in quantum gravity \cite{Isham:1992ms,Kuchar:1990vy,Brown:1994py,Husain:2011tk}. In particular in (\cite{Husain:2011tk}) a pressureless perfect fluid Lagrangian described by phase space fields $(T(x^{a}),P_{T}(x^{2}))$ was proposed, with it argued that the canonical gauge choice $T=t$ may be imposed prior to quantization. Notably, the classical equations of motion generally do not allow $T$ to be used as a global time variable due to the generic formation of caustics on surfaces of constant $T$ \cite{Babichev:2017lrx}. This suggests that if the flow of time $T$ in the quantum theory remains unimpeded, then quantum corrections to the classical equations of motion may become important near would-be caustic formation. An interesting illustration of this is how the classical Big Bang singularity of a dust-dominated universe can be replaced by an effective bounce when dust-time is used in a quantum cosmological description of the system \cite{Gielen:2020abd}. If degrees of freedom described by fields $(\phi^{I},\tilde{P}_{I})$ or $(T,\tilde{P}_{T})$ are to play the role of dark matter, then these corrections may have an experimental signature via how they affect the distribution and evolution of dark matter density.

\section{Additional terms in the gravitational action and coupling to matter}
\label{additional}
We now briefly consider additional terms in the gravitational action, the coupling of matter to gravitation, and the degree to which they affect the above results regarding how many local degrees of freedom are present in each. A straightforward extension of the models considered here would be to add a term with the following Lagrangian density:

\begin{align}
\tilde{\cal L} &= -\frac{\Lambda}{12}\tilde{\eps}^{\mu\nu\alpha\beta}\eps_{IJKL}D_{\mu}\phi^{I}D_{\nu}\phi^{J}D_{\alpha}\phi^{K}D_{\beta}\phi^{L}\label{cosmocon}
\end{align}
Where $\Lambda$ is a constant. It can be shown that this term modifies only the constraint (\ref{cicon}), contributing a term $(\Lambda/3!)\eps_{IJKL}\tilde{\eps}^{abc}e^{J}_{a}e^{K}_{b}e^{L}_{c}$, and that for general values of $(g_{+},g_{-})$ this term does not modify the character of the constraints and so does not affect the number of degrees of freedom of the theory. Indeed it can be shown that (\ref{cosmocon}) is invariant under the transformation (\ref{field_transformation}). In the cases $(g_{+}=1,g_{-}=0)$ and $(g_{+}=0,g_{-}=1)$ the effect of (\ref{cosmocon}) is that of a cosmological constant term.

We also now briefly consider other possible terms in the gravitational action. After integration by parts, the action (\ref{form_action}) can be expressed as a linear combination of integrated Lagrangian four-forms $\phi^{M}\phi_{M}\eps_{IJKL}R^{IJ}\wedge R^{KL}$ and $\phi_{J}\phi_{L}\eta_{IK}R^{IJ}\wedge R^{KL}$ \cite{Koivisto:2022uvd}. There is one additional,  independent four-form which is quadratic in $\phi^{I}$ and in $R^{IJ}$: $\phi^{K}\phi_{K} R_{IJ}\wedge R^{IJ}$. This term - and generally other terms involving $\phi_{I}\phi^{I}$ - may be excluded by the requirement that the gravitational action have a symmetry under `translations' $\phi^{I}\rightarrow \phi^{I}+p^{I}$ subject to $Dp^{I}=0$ (the action (\ref{form_action}) is manifestly invariant under this transformation).

The coupling of fields $(\phi^{I},\omega^{\pm IJ})$ to matter fields depends on the representation of the $SL(2,C)_{+} \times SL(2,C)_{-}$ gravitational symmetry that the matter field belongs to. For fields in the trivial representation, such as spacetime scalar fields $\varphi$ or one-forms $A_{\mu}$, coupling to gravity is expected to be entirely via the spacetime metric $g_{\mu\nu}=\eta_{IJ}D_{\mu}\phi^{I} D_{\nu}\phi^{J}$. In which case, in the canonical formalism, time derivatives of $\phi^{I}$ but not of $\omega^{\pm IJ}_{\mu}$ will appear in matter actions, leading to a modification of the definition of the momentum $\tilde{P}_{I}$ via additional terms appearing in (\ref{cicon}).
Additional couplings between gravity and matter fields are necessary when the fields belong to non-trivial representations of the gravitational symmetry group, such as left and right-handed fermions $\psi^{\pm}$. Here some of the gravitational gauge fields $\omega^{\pm IJ}_{\mu}$ must couple to $\psi^{\pm}$ to create covariant derivative terms for these fields; these couplings will introduce no new time derivatives of fields $(\phi^{I},\omega^{\pm IJ}_{\mu})$ but will result in additional terms in the constraints $\tilde{\cal G}^{\pm IJ}$.

\section{Conclusions}
\label{discussion}
We now briefly summarize the main results of the paper and discuss the potential for future work. In Sections \ref{introduction} to \ref{three_plus_one_lag} we introduced the models looked at in this paper and produced the Hamiltonian form of the models, focusing on an analysis of the propagation and nature of the phase space constraints present. The Lagrangian density of the class of models considered is:

\begin{align}
\tilde{\cal L}[\phi^{I},\omega^{\pm IJ}_{\mu}] &= \tilde{\eps}^{\mu\nu\alpha\beta}\eps_{IJKL}D_{\mu}\phi^{I}D_{\nu}\phi^{J}\bigg(g_{+}R^{KL}_{\ph{KL}\alpha\beta}(\omega^{+}) + g_{-}R^{KL}_{\ph{KL}\alpha\beta}(\omega^{-})\bigg) \label{gpgm}
\end{align}
It was found that in the general case $g_{+}\neq g_{-}$ that the theory possesses three complex degrees of freedom whereas for the particular case $g_{+}=g_{-}$ there are no local degrees of freedom. Furthermore, it was shown that the cases $(g_{+}=1,g_{-}=0)$ and $(g_{+}=0,g_{-}=1)$ correspond to an extension of General Relativity that includes solutions corresponding to an additional effective pressureless perfect fluid matter source. Interestingly, such a matter source has been of prior interest as a possible solution to the problem of time in quantum gravity \cite{Isham:1992ms,Brown:1994py,Husain:2011tm,Maniccia:2021skz,Maniccia:2022iqa}.
We find it encouraging that in the present models, the perfect fluid arises `naturally' from the theory and does not have to be independently posited.

The extension to General Relativity (\ref{lag_den_fin}) looks potentially promising: the requirement that a General Relativistic limit only occurs when the self-dual parts of the spatial pullback of the spin connection and its momenta represent the gravitational degrees of freedom implies that in this limit the gravitational Hamiltonian takes a simple, polynomial form. The presence of the new degree of freedom $\phi^{2}$ in this model produces a gravitational effect equivalent to that of a pressureless perfect fluid in the classical equations of motion. Given that an extremely wide array of cosmological and astrophysical data points towards the presence of an additional, unknown gravitational component that behaves as such a fluid on large scales \cite{Planck:2018vyg}, it is tempting to speculate whether the new degree of freedom may be responsible for at least some of this effect.

However, the model (\ref{khr_act_zero_one}) should be understood first as a quantum theory. Clearly, non-perturbative quantization of gravitational theories represents a considerable technical challenge so one approach is to first develop incremental results. 
As briefly discussed in Section \ref{phisqtime}, it is possible $\phi^{2}$ playing a putative role in determining a privileged and global time in quantum may have observable consequences. Furthermore, the case of Ashtekar's chiral formulation of gravity (whose canonical formulation arises from (\ref{lag_den_fin}) in the limit $\tilde{\Pi}\rightarrow 0$), it has been argued \cite{Magueijo:2010ba} that whatever the ultimate form that the quantized theory takes, it should possess a regime which is mappable to a classical cosmological background with metric perturbations describable in terms of the usual inflationary calculation of tensor vacuum quantum fluctuations. Given this assumption, it was found that the primordial spectrum of tensor modes of $+$ and $-$ chirality differed \cite{Magueijo:2010ba,Bethke:2011ru}, in contrast to the case of tensor modes in standard inflationary cosmology where gravity is described by metric General Relativity and no such effect exists. Remarkably, such effects in primordial tensor modes may be observable via their effect on the cross-correlation between CMB temperature and polarization fluctuations \cite{Contaldi:2008yz}. Given this, a first step that could be taken in the case of (\ref{lag_den_fin}) would be to allow for the effect of additional degrees of freedom $(\tilde{\Pi},\phi^{2})$ in the quantization of cosmological perturbations and see how the above picture is affected.

An alternative approach is to look to describe the behavior of the model as a quantum theory in situations of high symmetry. This has been carried out for Ashtekar's chiral formulation of gravity in the context of loop quantum cosmology \cite{Wilson-Ewing:2015lia,Wilson-Ewing:2015xaq,BenAchour:2014qca} and this approach could be generalized to the case of (\ref{lag_den_fin}).

Finally, it is interesting to note that despite the classical equivalence of parameterized field theory to the theory of matter fields propagating in fixed Minkowski space background, the Dirac quantization of the former theory faced technical obstacles \cite{Torre:1997zs}, the resolution of which required the use of techniques originally developed in the loop quantum gravity paradigm \cite{Varadarajan:2006am}.

A natural generalization of (\ref{gpgm}) would be the introduction of fields $(\psi_{+},\psi_{-})$ (potentially with non-trivial Lorentz index structure) such
that the hitherto constant $(g_{+},g_{-})$ are reflective of expectation values of these fields. The General-Relativistic limits $(g_{+}=1,g_{-}=0)$ and $(g_{+}=0,g_{-}=1)$
would then potentially arise from spontaneous symmetry breaking (with the action formally symmetric under the transformations (\ref{field_transformation}) and accompanying transformation of $(\psi_{+},\psi_{-})$) and with time variation of the new dynamical fields being of significance in the early universe \cite{NikjooRosaZlosnik23}. 

\section*{Acknowledgements}
We thank William Barker, Amel Durakovic, and Hans Westman for helpful discussions.
MJ and TZ are supported by the project No. 2021/43/P/ST2/02141 co-funded by the Polish National Science 
Centre and the European Union Framework Programme for Research and Innovation Horizon 2020 under the 
Marie Sk\l{}odowska-Curie grant agreement No. 945339.

\appendix
\section{Comparison to MacDowell-Mansouri gravity}
\label{macdowellmansouri}
In Section \ref{introduction} we discussed how the gravitational model considered in this paper could be recovered from the gauging of global symmetries present in a parameterized field theory whose solutions corresponded to fields propagating in a background Minkowski space. 
In the resulting gravitational theory, gravity is described by an $SO(1,3)_{C}$ connection $\omega^{I}_{\ph{I}J\mu}$ and a field $\phi^{I}$ in the group's fundamental representation. 

This bears an immediate resemblance to the description of gravity as a spontaneously broken gauge theory of the De Sitter ($SO(1,4)$) or anti-De Sitter groups ($SO(2,3)$). Initially proposed by MacDowell and Mansouri \cite{MacDowell:1977jt}, a manifestly locally $SO(1,4)$ or $SO(2,3)$- invariant action for gravity was proposed by Stelle and West \cite{Stelle:1979aj}, with the gravitational field described by an $SO(1,4)$ or $SO(2,3)$ connection $A^{A}_{\ph{A}B\mu}$ and a field $V^{A}$ in the group's fundamental representation.

The Stelle-West theory can additionally be considered to arise from the gauging of the parameterized field theory whose solutions correspond to fields propagating in a background De Sitter or anti-De Sitter space. For concreteness, we will focus on the De Sitter case, with the generalization to anti-De Sitter space straightforward. For this case, the non-gravitational parameterized field theory is obtained by replacing the fixed background De Sitter metric $\hat{\eta}_{\mu\nu}$ by a composite object $\eta_{AB}\partial_{\mu}X^{A}\partial_{\nu}X^{B}$ where $\eta_{AB}=\mathrm{diag}(-1,1,1,1,1)$ and the five dynamical fields $X^{A}$ are subject to the constraint $\eta_{AB}X^{A}X^{B}=L^{2}$. Then, the De Sitter metric is recovered if the equations of motion for matter fields and for $X^{A}$ allow for solutions where $X^{A}$ take the form of Minkowski coordinates in $\mathbb{R}^{(1,4)}$ and the condition $\eta_{AB}X^{A}X^{B}=L^{2}$ defines a De Sitter space submanifold. This submanifold can be covered with coordinates $x^{\mu}$, with the metric on this surface being the pullback of the metric $\eta_{AB}$ of $\mathbb{R}^{(1,4)}$ to this surface:  $\eta_{AB}\partial_{\mu}X^{A}\partial_{\nu}X^{B}$. A parameterized field theory can be constructed by replacing the fixed De Sitter background metric with $\eta_{AB}\partial_{\mu}X^{A}\partial_{\nu}X^{B}$ and adding a Lagrangian constraint $\eta_{AB}X^{A}X^{B}-L^{2}$ into the action. The resulting action has four-dimensional diffeomorphism symmetry as well as symmetry under the global De Sitter transformation

\begin{align}
X^{A} \rightarrow M^{A}_{\ph{A}B}X^{B}
\end{align}
with $M^{A}_{\ph{A}B} \in SO(1,4)$. This global symmetry can be promoted to a local one by the introduction of a De Sitter group connection $A^{A}_{\ph{A}B\mu}$ so that the covariant derivative ${\cal D}^{(A)}_{\mu}X^{A} = \partial_{\mu}X^{A}+A^{A}_{\ph{A}B\mu}X^{B}$ can be constructed. This suggests that the metric tensor that matter fields couple to should be identified with $\eta_{AB}{\cal D}^{(A)}X^{A}{\cal D}^{(A)}X^{B}$.

It is then necessary to introduce an action involving $A^{A}_{\ph{A}B\mu}$ so that equations of motion for this field don't produce artificial constraints on any matter fields it couples to. Stelle and West suggested the following action:

\begin{align}
S[A,X,\lambda] =  -\frac{\alpha}{4}\int d^{4}x \bigg(&\eps_{ABCDE}\eps^{\alpha\beta\mu\nu}X^{E}F^{AB}_{\ph{AB}\alpha\beta}F^{CD}_{\ph{CD}\mu\nu} - \lambda ( \eta_{AB}X^{A}X^{B}-L^{2})\bigg) \label{swact}
\end{align}
where $F^{AB}_{\ph{AB}\alpha\beta}$ is the curvature two-form of the connection $A^{A}_{\ph{A}B\mu}$ and the Lagrangian constraint on the norm of $X^{A}$ is retained from the non-gravitational theory. The $\lambda$ equation of motion $\eta_{AB}X^{A}X^{B}-L^{2}=0$ enforces a non-vanishing vacuum expectation value of $X^{A}$, spontaneously breaking the $SO(1,4)$ symmetry down to $SO(1,3)$ It is useful to gauge-fix at the level of the action. Choosing a gauge where $X^{A} \overset{*}{=} L\delta^{A}_{-1}$ (with $\eta_{-1-1}=1$ and indices $I,J,\dots$ used to label remaining indices), a useful decomposition of the connection $A^{A}_{\ph{A}B\mu}$ is:

\begin{align}
A^{AB}_{\ph{AB}\mu} \overset{*}{=}
\begin{pmatrix}
   0& \frac{1}{L}e^{I}_{\mu} \\
    -\frac{1}{L}e^{I}_{\mu} & \omega^{IJ}_{\ph{IJ}\mu}
\end{pmatrix} \label{so14con}
\end{align}
From which it follows that:

\begin{align}
g_{\mu\nu} &\overset{*}{=} \eta_{IJ}e^{I}_{\mu}e^{J}_{\nu}\\
F^{IJ}_{\ph{IJ}\mu\nu} &\overset{*}{=} R^{IJ}_{\ph{IJ}\mu\nu}(\omega) - \frac{2}{L^{2}}e^{I}_{[\mu}e^{J}_{\nu]}  \label{fij}\\
F^{-1 I}_{\ph{-1I}\mu\nu} &\overset{*}{=}  \frac{2}{L}{\cal D}^{(\omega)}_{[\mu}e^{I}_{\nu]} \label{f4i}
\end{align}
where $R^{IJ}_{\ph{IJ}\mu\nu}(\omega)$ is to be identified with the curvature of the $SO(1,3)$ connection $\omega^{IJ}_{\mu}$ with ${\cal D}^{(\omega)}_{\mu}$ being the associated covariant derivative. Inserting (\ref{fij}) and (\ref{f4i}) in (\ref{swact}), three independent contributions to the action are recovered: a term quadratic in $e^{I}_{\mu}$ and linear in $R^{IJ}_{\mu\nu}$ which is the Palatini action of Einstein-Cartan gravity, a term quartic in $e^{I}_{\mu}$ which yields a positive cosmological constant, and a term quadratic in $R^{IJ}_{\ph{IJ}\mu\nu}$ which is a topological term. Indeed, the gauged-fixed action corresponds to the action considered by MacDowell and Mansouri.

An independent interpretation of the recovery of gravity as a spontaneously broken gauge theory is via the framework of Cartan geometry \cite{Wise:2006sm,Westman:2014yca}, where the connection (\ref{so14con}) can be interpreted as a prescription for `rolling without slipping' of a model De Sitter space on a local patch of physical spacetime, taken to be a four-dimensional surface embedded in $\mathbb{R}^{(1,4)}$. 
By comparison, for the $SO(1,3)_{C}$ model with a General Relativistic limit (without loss of generality we choose $\beta=i$), consider a patch of spacetime where $\eta_{IJ}\phi^{I}\phi^{J}<0$. It's helpful to choose a partial gauge fixing where $\phi^{I} \overset{*}{=} T(x^{\mu})\delta^{I}_{0}$ (where $\eta_{00}=-1$). Then from the definition $e_{\mu}^{I}\equiv {\cal D}_{\mu}X^{I}$ we have $e_{\mu}^{0} \overset{*}{=}  \partial_{\mu}T$, $e_{\mu}^{i} \overset{*}{=} {\cal A}^{i}_{\ph{i}0\mu}$ it can then be shown from the field equations that if $T$ is adopted as a time coordinate then the pullback of $\omega^{IJ}_{\mu}$ to a patch of constant $T$ takes the form:

\begin{align}
\omega^{IJ}_{\ph{IJ}a} &\overset{*}{=}  \begin{pmatrix} 
0 & \frac{1}{T}e_{a}^{i} \\
-\frac{1}{T}e^{i}_{a} & \Gamma^{ij}_{\ph{ij}a} -i\eps^{ij}_{\ph{ij}k}(\Gamma^{0k}_{\ph{0k}a}-\frac{1}{T}e_{a}^{k})\\
\end{pmatrix}
\end{align}
where $\Gamma^{ij}_{\ph{ij}a}$ is the torsion-free spin connection associated with the spatial triad field $e^{i}_{a}$ and $\Gamma^{0k}_{\ph{0k}a}$ is proportional the extrinsic curvature of the patch. This illustrates how $\omega^{IJ}_{\ph{IJ}a}$ encodes information about distances on the patch at constant $T$ (via $e^{i}_{a}$), the intrinsic curvature of the patch (via $\Gamma^{ij}_{\ph{ij}a}$), and its extrinsic curvature in spacetime (via $\Gamma^{0k}_{\ph{0k}a}$). The first two of these parts have much in common with the Cartan geometrodynamical approach to the Hamiltonian formulation of gravity \cite{Gielen:2011mk}, whereas the incorporation of information about the time evolution of space via complexification of $\omega^{IJ}_{\ph{IJ}a}$ is to our knowledge novel.

\section{Useful functional derivatives}
\label{functional_derivatives}
Useful non-zero functional derivatives are as follows:

\begin{align}
\frac{\delta \beta^{+IJ}_{a}(x)}{\delta \beta^{+KL}_{b}(x')} &= \frac{1}{2}\bigg(\delta^{[I}_{K}\delta^{J]}_{L}-\frac{i}{2}\eps_{KL}^{\ph{KL}IJ}\bigg)\delta^{b}_{a} \delta(x-x')\\
\frac{\delta \tilde{P}^{+a}_{IJ}(x)}{\delta \tilde{P}^{+b}_{KL}(x')} &=  \frac{1}{2}\bigg(\delta^{[K}_{I}\delta^{L]}_{J}-\frac{i}{2}\eps_{IJ}^{\ph{IJ}KL}\bigg)\delta^{a}_{b}\delta(x-x')\\
\frac{\delta \beta^{-IJ}_{a}(x)}{\delta \beta^{-KL}_{b}(x')} &= \frac{1}{2}\bigg(\delta^{[I}_{K}\delta^{J]}_{L}+\frac{i}{2}\eps_{KL}^{\ph{KL}IJ}\bigg)\delta^{b}_{a} \delta(x-x')\\
\frac{\delta \tilde{P}^{-a}_{IJ}(x)}{\delta \tilde{P}^{-b}_{KL}(x')} &=  \frac{1}{2}\bigg(\delta^{[K}_{I}\delta^{L]}_{J}+\frac{i}{2}\eps_{IJ}^{\ph{IJ}KL}\bigg)\delta^{a}_{b}\delta(x-x')\\
\frac{\delta \tilde{P}_{I}(x)}{\delta \tilde{P}_{J}(x')} &= \delta^{J}_{I}\delta(x-x')\\
\frac{\delta \phi^{I}(x)}{\delta \phi^{J}(x')} &= \delta^{I}_{J}\delta(x-x')
\end{align}
\section{Detailed example of the evaluation of the Poisson bracket between two constraints}
\label{poisson_example}
Making use of the functional derivatives defined in the previous section and the definition of the Poisson bracket (\ref{poisson_bracket}), as an illustrative example we consider a detailed calculation of the Poisson bracket $\{\tilde{C}^{-a}_{IJ}[A_{a}^{IJ}],\tilde{C}_{K}[V^{K}]\}$. To do so, we first calculate the functional derivative of each smeared constraint as follows:

\subsection{Functional derivatives of $\tilde{C}^{-a}_{IJ}[A_{a}^{IJ}]$}
\label{CAB_fds}
Given a test function $A_{a}^{IJ} = A_{a}^{-IJ}$ (i.e. a Lorentz tensor which depends on of $x^{a}$ and is considered independent of phase space fields), we can consider the following smeared constraint:

\begin{align}
\tilde{C}^{-a}_{IJ}[A_{a}^{IJ}] &= \int d^{3}x	A_{c}^{IJ}\bigg(\tilde{P}_{IJ}^{-c} - 2g_{-} \bigg[\eps_{IJKL}\tilde{\vareps}^{abc}e^{K}_{a}e^{L}_{b}\bigg]^{-}\bigg)
\end{align}
where recall that $e^{I}_{a}=\partial_{a}\phi^{I}+\beta^{I}_{\ph{I}Ja}\phi^{J}$. Using the results of Section \ref{functional_derivatives} we have:

\begin{align}
	\frac{\delta\tilde{C}^{-a}_{KL}[A_{a}^{KL}]}{\delta \tilde{P}^{+d}_{IJ}} &= 0
\end{align}
\begin{align}
	\frac{\delta\tilde{C}^{-a}_{KL}[A_{a}^{KL}]}{\delta \beta^{+IJ}_{d}} &= 4g_{-}\bigg[A_{c}^{-KL} \eps_{MKL[I}\tilde{\vareps}^{dbc}\phi_{J]}e_{b}^{M}\bigg]^{+} 
\end{align}

\begin{align}
	\frac{\delta\tilde{C}^{-a}_{KL}[A_{a}^{KL}]}{\delta \tilde{P}^{-d}_{IJ}} &= A^{IJ}_{d}
\end{align}
\begin{align}
	\frac{\delta\tilde{C}^{-a}_{KL}[A_{a}^{KL}]}{\delta \beta^{-IJ}_{d}} &= 4g_{-}\bigg[A_{c}^{-KL} \eps_{MKL[I}\tilde{\vareps}^{dbc}\phi_{J]}e^{M}_{b}\bigg]^{-}
\end{align}

\begin{align}
	\frac{\delta\tilde{C}^{-a}_{KL}[A_{a}^{KL}]}{\delta \phi^{I}} &=4g_{-}D^{(\beta)}_{a}\bigg(A_{c}^{-KL} \eps_{IKLM}\tilde{\vareps}^{abc}e^{M}_{b}\bigg) 
\end{align}
\begin{align}
	\frac{\delta\tilde{C}^{-a}_{KL}[A_{a}^{KL}]}{\delta \tilde{P}_{I}} &=0
\end{align}

\subsection{Functional derivatives of $\tilde{C}_{I}[A^{I}]$}
\label{CA_fds}
Similarly for the constraint $\tilde{C}_{I}$ we may consider a test function $A^{I}$ and smear the constraint as follows:

\begin{align}
\tilde{C}_{I}[A^{I}] &= \int d^{3}x A^{I}\bigg[\tilde{P}_{I} - 2g_{+} \eps_{IJKL}\tilde{\vareps}^{abc}e^{J}_{a}R_{bc}^{+KL} -2g_{-} \eps_{IJKL}\tilde{\vareps}^{abc}e^{J}_{a}R^{-KL}_{bc}\bigg]
\end{align}
Again using the results of Section \ref{functional_derivatives} we have:
\begin{align}
\frac{\delta\tilde{C}_{K}[A^{K}] }{\delta \tilde{P}^{+d}_{IJ}} &=0
\end{align}
\begin{align}
\frac{\delta\tilde{C}_{K}[A^{K}] }{\delta \beta^{+IJ}_{d}} &=  \bigg[- 2A^{M}\eps_{MKL[I}\tilde{\vareps}^{dbc}\phi_{J]}(g_{+} R^{+KL}_{bc}+g_{-} R^{-KL}_{bc}) -4 g_{+}\tilde{\vareps}^{bad}D^{+}_{b}\bigg(\eps_{IJKL}A^{K}e_{a}^{L}\bigg)\bigg]^{+}
\end{align}
\begin{align}
\frac{\delta\tilde{C}_{K}[A^{K}] }{\delta \tilde{P}^{-d}_{IJ}} &=0
\end{align}
\begin{align}
\frac{\delta\tilde{C}_{K}[A^{K}] }{\delta \beta^{-IJ}_{d}} &= \bigg[- 2A^{M}\eps_{MKL[I}\tilde{\vareps}^{dbc}\phi_{J]}(g_{+} R^{+KL}_{bc}+g_{-} R^{-KL}_{bc}) -4 g_{-}\tilde{\vareps}^{bad}D^{-}_{b}\bigg(\eps_{IJKL}A^{K}e_{a}^{L}\bigg)\bigg]^{-}
\end{align}

\begin{align}
\frac{\delta\tilde{C}_{K}[A^{K}] }{\delta \phi^{I}} &=2 g_{+}D^{(\beta)}_{a}\bigg(A^{M}\eps_{MIKL}\tilde{\vareps}^{abc}R^{+KL}_{bc}\bigg) +2 g_{-}D^{(\beta)}_{a}\bigg(A^{L}\eps_{LIJK}\tilde{\vareps}^{abc}R^{-JK}_{bc}\bigg) 
\end{align}
\begin{align}
\frac{\delta\tilde{C}_{J}[A^{J}] }{\delta \tilde{P}_{I}} &= A^{I}
\end{align}
\subsection{Poisson bracket $\{\tilde{C}^{-a}_{IJ}[A_{a}^{IJ}],\tilde{C}_{K}[V^{K}]\}$}
Applying the results from Sections \ref{CAB_fds} and Sections \ref{CA_fds} we have that:
\begin{align}
  \{\tilde{C}^{-c}_{IJ}[A_{c}^{IJ}],\tilde{C}_{K}[V^{K}]\}  &\overset{b}{=} \int d^{3}x A^{IJ}_{d}V^{K}\bigg[ 2(g_{+}-g_{-})\tilde{\vareps}^{dbc}\eps_{KLM[I}\phi_{J]}R_{bc}^{+LM}+ g_{-}\tilde{\vareps}^{dbc}\phi_{K}\eps_{IJMN}R^{-MN}_{bc}\bigg]^{-}
\end{align}
where $\overset{b}{=}$ denotes equality up to a total derivative (and hence boundary) term.

%\begin{align}
%\{\tilde{C}^{-c}_{CD}[A_{c}^{CD}],\tilde{C}_{E}[V^{E}]\} &=  -2\int d^{3}x A_{d}^{IJ} V^{A}\eps_{ACDI}\tilde{\vareps}^{dbc}\phi_{J}R^{CD}_{bc}
%\end{align}

\section{Interpretation of the Lagrange multiplier $V^{I}$}
\label{vint}
From Hamilton's equations we have that

\begin{align}
\dot{\phi}^{I} &= V^{I} - (\Omega^{+IJ} + \Omega^{-IJ})\phi_{J}
\end{align}
Therefore, using the results of Section \ref{spacetime_structure} we have that:

\begin{align}
V^{I} = D_{t}^{(\omega)}\phi^{I} &=  e^{I}_{t} = NN^{I} + N^{a}e_{a}^{I}\label{vnn}
\end{align}
and hence $V^{I}$ is straightforwardly related to parts of the spacetime metric structure. 

\section{Some useful results}
\label{usefulresults}
\subsection{Self and anti-self dual parts of the Lorentz tensor $Y^{de}_{IJPR}$} 
Calculation shows self and anti-self dual parts of the Lorentz tensor $Y^{de}_{IJKL}$ are given by:

\begin{align}
Y^{de}_{[IJ]^{(1)\pm}[KL]^{(2)\pm}} &=\tilde{\vareps}^{dbe}e^{M}_{b}
\eps_{MIJ[K}\phi_{L]}+(\pm)^{(1)} (\pm )^{(2)}\tilde{\vareps}^{dbe}\phi^{M}
e_{b[I}\eps_{J]MKL}\nn\\
& +2i\tilde{\vareps}^{dbe}e^{M}_{b}((\pm)^{1}
\eta_{M[I}\eta_{J][K}\phi_{L]}+(\pm )^{(2)}\phi_{[I}\eta_{J][L}\eta_{K]M}) \nn\\
&-(\pm )^{(2)}\tilde{\vareps}^{dbe}e^{M}_{b}\phi_{M}\eta_{K[I}\eta_{J]L} \label{Yplusmin}
\end{align}
\subsection{Development of the constraint $\tilde{C}^{I}$ in the case $(g_{+}=1,g_{-}=0)$}
Using (\ref{vnn}) we see that the constraint $\tilde{C}_{I}$ contributes to the Hamiltonian density via

\begin{align}
    V^{I}\tilde{C}_{I}   &=
     (NN^{I}+N^{a}e_{a}^{I})\bigg(\tilde{P}_{I} - 2\eps_{IJKL}\tilde{\vareps}^{abc}e^{J}_{a}R_{bc}^{+KL}\bigg)
\end{align}
Furthermore, when the constraints hold we have $\tilde{P}_{IJ}^{+ c}=-2\big[\eps_{IJKL}\tilde{\vareps}^{abc}e^{K}_{a}e^{L}_{b}\big]^{+}$. We can therefore recover the following useful results. Firstly, the product of the determinant of $q_{ab}$ and its matrix inverse can be expressed in terms of momenta $\tilde{P}^{+c}_{IJ}$ as:
\begin{align}
      -16q q^{cd} &=  \tilde{P}^{+c}_{IJ} \tilde{P}^{+dIJ} \label{momdet}
\end{align}
Furthermore we can express individual parts of $\tilde{C}^{I}$ as:

\begin{align}
\eps_{IJKL}\tilde{\vareps}^{abc}e_{d}^{I}e_{a}^{J}R^{+KL}_{bc} &= -\frac{1}{2}\tilde{P}^{+e}_{IJ}R^{+IJ}_{de}\\
-2\sqrt{q}N^{I}  \eps_{IJKL}e_{a}^{J}\tilde{\eps}^{abc}R_{bc}^{+KL} &= -\frac{1}{4}\tilde{P}^{+b}_{IK}\tilde{P}^{+cK}_{\ph{+cK}J}R^{+IJ}_{\ph{+IJ}bc}\\
 N^{I}\tilde{P}_{I} &= \xi \tilde{\Pi}\sqrt{-\phi^{2}+ \frac{1}{4}q^{ab}\partial_{a}\phi^{2}\partial_{b}\phi^{2}}\\
e_{a}^{I}\tilde{P}_{I} &= \frac{1}{2}\tilde{\Pi}\partial_{a}\phi^{2}
\end{align}
where we have used the result that $\tilde{P}^{I}\propto \phi^{I}$ in the cases where $(g_{+}=1,g_{-}=0)$ and $(g_{-}=0,g_{+}=1)$.

\bibliographystyle{hunsrt}
\bibliography{references}
\end{document}